\documentclass[aps,twocolumn]{revtex4}
\usepackage[colorlinks, linkcolor=red,anchorcolor=green,citecolor=blue]{hyperref}
\usepackage{graphicx}
\usepackage{bm}
\usepackage{amsmath}
\usepackage{color}
\usepackage{booktabs}
\usepackage{multirow}
\usepackage{diagbox}
\usepackage{xcolor, color, framed, ulem}
\graphicspath{{./fig/}}

\begin{document}

\title{Mass spectra and decays of open-heavy tetraquark states}
\author{Tao Guo}
\author{Jianing Li}
\author{Jiaxing Zhao}
\author{Lianyi He}
\affiliation{Physics Department, Tsinghua University, Beijing 100084, China\\}

\date{\today}%

\begin{abstract}
Open-heavy tetraquark states, especially those contain four different quarks have drawn much attention in both theoretical and experimental fields. In the framework of the improved chromomagnetic interaction (ICMI) model, we complete a systematic study on the mass spectra and possible strong decay channels of the $S$-wave open-heavy tetraquark states, $qq\bar{q}\bar{Q}$ ($q=u,d,s$ and $Q=c,b$), with different quantum number $J^P=0^+$, $1^+$, and $2^+$. The parameters in the ICMI model are extracted from the conventional hadron spectra and used directly to predict the mass of tetraquark states. Several compact bound states and narrow resonances are found in both charm-strange and bottom-strange tetraquark sectors, most of them as a product of the strong coupling between the different channels. 
Our results show the recently discovered four different flavors tetraquark candidates $X_0(2900)$ is probably compact $ud\bar{s}\bar{c}$ state with quantum number $J^P=0^+$. The predictions about $X_0(2900)$ and its partners are expected to be better checked with other theories and future experiments.
\end{abstract}

%\keywords{Suggested keywords} %Use showkeys class option if keyword

\maketitle
%\tableofcontents

\section{Introduction}
The quark model proposed by M. Gell-Mann and G. Zweig independently~\cite{GellMann:1964nj}, is not only a successful classification scheme for hadrons (only spatial ground states) but also indicates the existence of the ”elementary” particles which are named quarks, and the color degree of freedom of  ”elementary” particles. Although the quark model can be derived from the theory of quantum chromodynamics (QCD), the structure of hadrons is much more complicated than the quark model allows.
Any color-neutral configurations are approved in the QCD theory, such as the Glueballs that containing only valence gluons~\cite{Mathieu:2008me,Celi:2013gma}, multi-quark states (e.g. tetraquarks and pentaquarks)~\cite{Esposito:2016noz,Karliner:2017qhf}, hybrids which is composed of quarks and gluons~\cite{Meyer:2015eta,Chanowitz:1982qj}, and hadronic molecules~\cite{DeRujula:1976zlg,Guo:2017jvc}. So far, there are a lot of candidates for exotic hadrons in the light quark sector, such as $a_ 0(980)$, $f_0(1370)$, $\Lambda(1405)$, and so on. The discovery of charmonium (bottomonium)-like exotic states which names XYZ mesons~\cite{Richard:2016eis,Ali:2017jda,Liu:2019zoy,Brambilla:2019esw}, fully-heavy tetraquark state~\cite{LHCb:2020bwg,Karliner:2016zzc,Lu:2020cns,Liu:2019zuc,Wang:2019rdo,Zhu:2020xni,Bicudo:2015vta,Chen:2016jxd,Zhao:2020nwy}, and pentaquarks~\cite{LHCb:2015yax,LHCb:2019kea,Liu:2019zoy} attract much attention of experimenters and theorists. Besides, the X(5568) is believed to be the first exotic candidate with four different flavors of $u\bar ds\bar b$ or $\bar uds\bar b$ tetraquark states observed experimentally~\cite{D0:2016mwd,D0:2017qqm}. However, its existence was not confirmed by LHCb~\cite{LHCb:2016dxl}, CMS~\cite{CMS:2017hfy}, CDF~\cite{CDF:2017dwr}, and ATLAS~\cite{ATLAS:2018udc} collaborations.

Recently, the LHCb collaboration observes an enhancement on the $D^-K^+$ invariant mass distribution in the decay channel of $B^+\to D^+D^-K^+$~\cite{LHCb:2020bls,LHCb:2020pxc}. The best fit gives two possible states, $X_0(2900)$ with $J^{P}=0^+$, mass $M=2.866\pm0.007\pm0.002~\text{GeV}$, width $\Gamma=57\pm12\pm4~\text{MeV}$ and $X_1(2900)$ with $J^{P}=1^-$, mass $M= 2.904\pm0.005\pm0.001~\text{GeV}$, width $\Gamma=110\pm11\pm4 ~\text{MeV}$. 
From the decay products, one can infer the quark composition of $X_{0,1}(2900)$ is $ud\bar s\bar c$. 
If this state is confirmed, it will really be the first exotic candidate with four different flavors. A wave of research was aroused in the hadron community again. 
On the theoretical side, there are many controversies about the inner structures and properties of $X_{0,1}(2900)$.
One explanation for the enhancement of the $D^-K^+$ invariant mass distribution is the rescattering effects~\cite{Molina:2010tx,Chen:2020eyu,Guo:2019twa}. Molecular interpretation of $X_0(2900)$ is supported by the QCD sum rules~\cite{Chen:2020aos,Agaev:2020nrc}, one-boson-exchange models~\cite{Liu:2020nil,He:2020btl}, and effective field theories~\cite{Molina:2020hde,Hu:2020mxp}. Those works show $X_0(2900)$ is the bound state of $\bar D^*K^*$, while $X_1(2900)$ seems not a molecular state~\cite{Liu:2020nil}. In addition, the results based on the QCD sum rules~\cite{Wang:2020xyc,Zhang:2020oze,Mutuk:2020igv}, relativistic/non-relativistic quark models~\cite{Wang:2020prk,Yang:2021izl}, and chromomagnetic interaction models~\cite{He:2020jna} indicate both $X_0(2900)$ and $X_1(2900)$ are compact tetraquark states.

Chromomagnetic interaction (CMI) mode which is proposed to explain the mass splittings of the conventional hadrons has been extended to study the tetraquark and pentaquark states~\cite{Silvestre-Brac:1992kaa,Buccella:2006fn,Hogaasen:2004pm,Anselmino:1992vg,Lebed:2015tna,Maiani:2004uc,Rossi:2016szw,Luo:2017eub,Wu:2017weo,Wu:2018xdi,Cheng:2020nho}(see Ref.~\cite{Liu:2019zoy} for a review). In the CMI model, one does not need to solve the few-body bound state problem directly since the contributions from the spatial part of the total wave function have been encoded in the effective parameters. The mass splittings of different hadron states with the same quark components can be given by the color-spin interaction term and the hadron masses can be estimated by adding effective quark masses. In the conventional CMI model, chromoelectric effects from the color interaction are absorbed implicitly in effective quark masses. However, it is difficult to account for the two-body chromoelectric effects in all mesons and baryons by the effective quark masses~\cite{Hogaasen:2013nca,Weng:2018mmf,An:2021vwi,Weng:2019ynv}, especially, the multi-quark systems which have two or more color configurations. Besides, the CMI model with only the short-range chromomagnetic interaction is not suitable/enough to be used to interpret the open-heavy tetraquark states which have a large size, While the chromoelectric effect is needed. In this paper, we want to use the improved CMI model to investigate the mass spectra of open-heavy tetraquark states and find possible structures and decay channels of newly observed states $X_{0}(2900)$ and their partners.   

This article is organized as follows. In Sec. II, we introduce the framework of the improved CMI model. The parameters such as effective masses and coupling strengths are shown in Sec. III. The mass spectra and decay channels of various open-heavy tetraquark states are shown and analyzed in the sametime. In the last section, we will give some discussions and a summary of this work.

\section{ The theoretical approach}
In ICMI model, the mass spectrum of tetraquark system is given by solving the expectation value of the Hamiltonian~\cite{Hogaasen:2013nca,Weng:2018mmf}:
 \begin{equation}\label{Ha}
H=\sum\limits_{i=1}^4m_i+H_{cm} + H_{ce},
 \end{equation}
where $m_i$ is the effective mass of the $i$-th constituent quark and $H_{cm}$ is the chromomagnetic interaction term arising from the one-gluon exchange in the MIT bag model.
Besides, $H_{ce}$ is expressed as the chromoelectric interaction term.
Here, the chromomagnetic interaction Hamiltonian acting on the color and spin space of quarks can read~\cite{DeRujula:1975qlm,Maiani:2004vq,Cui:2005az,Hogaasen:2005jv,Guo:2011gu,Kim:2014ywa}:
\begin{equation}\label{cm}
H_{cm}=-\sum\limits_{i<j}v_{ij}{\lambda}_{i}^{c}\cdot{\lambda}_{j}^{c}
\vec{\sigma}_{i}\cdot\vec{\sigma}_{j},
\end{equation}
and the chromoelectric interaction Hamiltonian is~\cite{Oka:1989ud,Hogaasen:2013nca}:
\begin{equation}\label{ce}
H_{ce}=-\sum\limits_{i<j}c_{ij}{\lambda}_{i}^{c}\cdot{\lambda}_{j}^{c}.
\end{equation}
In Eq.(\ref{cm}) and Eq.(\ref{ce}), ${\lambda}_{i}^{c}$ ($c=1, ..., 8$) and $\vec{\sigma}_i$ are the Gell-Mann matrix and the Pauli matrix for the $i$-th quark, respectively.
Furthermore, if the subscript $i$ or $j$ indicates an antiquark, $\lambda_i^c$ should be replaced by $-\lambda_i^{c*}$. The values of the interactional strength parameters $v_{ij}$ and $c_{ij}$ are related to the constituent quark masses and the structural characteristics of the spatial wave function of the tetraquark systems.
For $S$-wave states, generally, this approach of the improved chromomagnetic interaction model has achieved great success in describing the mass spectra of mesons and baryons~\cite{Weng:2018mmf}.
Therefore, for the sake of simplicity, we can obtain these parameters, $m_i$, $v_{ij}$ and $c_{ij}$, by fitting the meson and baryon spectra.

Without loss of generality and substituting Eq.(\ref{cm}) and Eq.(\ref{ce}) into Eq.(\ref{Ha}), the Hamiltonian can be simplified to
 \begin{equation}\label{Ha1}
 H=H_0+H_{cm},
 \end{equation}
with
 \begin{equation}\label{H0}
H_0=-\frac{3}{16}\sum\limits_{i<j}m_{ij}{\lambda}_{i}^{c}\cdot{\lambda}_{j}^{c},
 \end{equation}
where the newly introduced parameter $m_{ij}=m_i+m_j+16c_{ij}/3$~\cite{Weng:2018mmf}, $m_{ij}$ is related to the effective mass $m_i$ of the constituent quark and the coupling strength coefficient $c_{ij}$ of the chromoelectric interaction.
%\footnote{In the derivation process of the formula (\ref{H0}), we applied the equation $\sum\limits_{i<j}(m_i+m_j)\tilde{\lambda}_{i}^{c}\cdot\tilde{\lambda}_{j}^{c}=(\sum\limits_{i}m_i\tilde{\lambda}_{i}^{c})\cdot(\sum\limits_{i}\tilde{\lambda}_{i}^{c})-\frac{16}{3}\sum\limits_{i}m_i$, where the color operator $\sum\limits_{i}\tilde{\lambda}_{i}^{c}$ vanish to act on any colorless hadron state.}.

Physically, the mass spectrum $M$ of tetraquark systems can be solved by the analytical formula
 \begin{equation}\label{m}
 M=\langle H_0\rangle+\langle H_{cm}\rangle,
 \end{equation}
where $\langle H_{cm} \rangle$ is the expectation value of the chromomagnetic interaction with proper account for the flavor symmetry breaking.
And  $\langle H_{0} \rangle$ is the expectation value of the Eq.(\ref{H0}), which includes the chromoelectric interaction and the effective quark mass.

So far, the hadrons observed in experiments are all color singlets, which are theoretically described as the phenomenon of color confinement. Considering $SU(3)$ symmetry, the colors of quark and antiquark can be defined as $3$ and $\bar{3}$, respectively. For tetraquark state, there are two kinds of decompositions in color space,
\begin{eqnarray}
3\otimes 3 \otimes \bar 3 \otimes \bar 3 &=& (3\otimes \bar 3) \otimes (3 \otimes \bar 3)\nonumber\\
&=&8\otimes8  \oplus 1\otimes1 \oplus  1\otimes8 \oplus 8\otimes1, \nonumber\\
3\otimes 3 \otimes \bar 3 \otimes \bar 3&=& (3\otimes 3) \otimes (\bar 3 \otimes \bar 3)\nonumber\\
&=& \bar 3 \otimes 3 \oplus  6 \otimes \bar 6 \oplus  \bar 3 \otimes \bar 6 \oplus   6 \otimes 3.
\end{eqnarray}
They are corresponding to two different configures in color space, diquark anti-diquark configure $|(q_1q_2)(\bar q_3\bar{q}_4)\rangle$ and meson-meson configure $|(q_1\bar{q}_3)(q_2\bar{q}_4)\rangle$ (or $|(q_1\bar{q}_4)(q_2\bar{q}_3)\rangle$). We can see there are two color-singlet states obtained from the first and
second terms on the right-hand side of second and forth line. They can be expressed as,
\begin{eqnarray}
|(q_1\bar q_3)^1(q_2\bar q_4)^1\rangle^1, ~ |(q_1\bar q_3)^8(q_2\bar q_4)^8\rangle^1,
\end{eqnarray}
in meson-meson configure. Similar for $[(q_1\bar{q}_4)(q_2\bar{q}_3)]$ system. And 
\begin{eqnarray}
|(q_1q_2)^{\bar 3}(\bar q_3\bar q_4)^{3}\rangle^1, ~|(q_1q_2)^6(\bar q_3\bar q_4)^{\bar 6}\rangle^1,
\end{eqnarray}
in diquark anti-diquark configure. The superscript denote the color of the subsystems $q_1\bar q_3$ and $q_2\bar q_4$
(or $q_1q_2$ and $\bar q_3\bar q_4$) and the whole tetraquark system.
These two set of color-singlet states are connected with each other via a linear transformation, \begin{eqnarray}
|(q_1\bar q_3)^1(q_2\bar q_4)^1\rangle^1&=&\sqrt{{1\over 3}}|(q_1q_2)^{\bar 3}(\bar q_3\bar q_4)^{3}\rangle^1\nonumber\\
&+&\sqrt{{2\over 3}}|(q_1q_2)^6(\bar q_3\bar q_4)^{\bar 6}\rangle^1. \nonumber\\
|(q_1\bar q_3)^8(q_2\bar q_4)^8\rangle^1&=&-\sqrt{{2\over 3}}|(q_1q_2)^{\bar 3}(\bar q_3\bar q_4)^{3}\rangle^1\nonumber\\
&+&\sqrt{{1\over 3}}|(q_1q_2)^6(\bar q_3\bar q_4)^{\bar 6}\rangle^1.
\label{13-12}
\end{eqnarray}
Therefore, the corresponding matrix elements can be obtained by the Casimir operator acting on the above color basis vectors. In this work, for better studying the decay characteristics of the tetraquark states, it is convenient to choose the basis $|(q_1\bar{q}_3)(q_2\bar{q}_4)\rangle$ or $|(q_1\bar{q}_4)(q_2\bar{q}_3)\rangle$.

Similarly, the direct product decomposition in spin-space gives,
\begin{eqnarray}
2\otimes 2 \otimes 2 \otimes 2 = 5 \oplus 3 \oplus 3 \oplus  3 \oplus 1 \oplus 1.
\end{eqnarray}
So, the total spin of the possible $S$-wave tetraquark states can take $0,1,2$.
%We can label the spin wavefunction of tetraquark system into:
There are two spin zero states,
\begin{eqnarray}
|(q_1\bar q_3)_0(q_2\bar q_4)_0\rangle_0, ~ |(q_1\bar q_3)_1(q_2\bar q_4)_1\rangle_0.
\end{eqnarray}
Three spin 1 states,
\begin{eqnarray}
|(q_1\bar q_3)_0(q_2\bar q_4)_1\rangle_1, ~ |(q_1\bar q_3)_1(q_2\bar q_4)_0\rangle_1, ~
|(q_1\bar q_3)_1(q_2\bar q_4)_1\rangle_1. \nonumber\\
\end{eqnarray}
And one spin 2 state,
\begin{eqnarray}
|(q_1\bar q_3)_1(q_2\bar q_4)_1\rangle_2.
\end{eqnarray}
where the subscripts denote the spin of the subsystems $q_1\bar q_3$ and $q_2\bar q_4$ and the whole tetraquark system.

Further, in order to solve the mass spectrum of the tetraquark state, we should get the chromomagnetic Hamiltonian matrix Eq.(\ref{cm}) and chromoelectric related Hamiltonian matrix Eq.(\ref{H0}) firstly.
Considering a certain symmetry relationship, therefore, we can construct all possible color-spin wave function bases in the tetraquark systems for these given quantum numbers $J^P=0^{+}$, $1^{+}$, and $2^{+}$.

For the scalar tetraquark systems with $J^P=0^{+}$, the color-spin basis vectors can be built in meson-meson configure and labeled as $\alpha$, 
\begin{equation}
\label{1324basis0}
\left.
\begin{array}{ll}
\hspace{-2mm}
\alpha_1\equiv |(q_1\bar{q}_3)^1_0\otimes(q_2\bar{q}_4)^1_0\rangle, & \alpha_2\equiv |(q_1\bar{q}_3)^1_1\otimes(q_2\bar{q}_4)^1_1\rangle,\\[0.5em]
\hspace{-2mm}\alpha_3\equiv |(q_1\bar{q}_3)^8_0\otimes(q_2\bar{q}_4)^8_0\rangle, & \alpha_4\equiv |(q_1\bar{q}_3)^8_1\otimes(q_2\bar{q}_4)^8_1\rangle,
\end{array}
\right.
\end{equation}
where the superscript and subscripts denote the color and spin of the subsystems $q_1\bar{q}_3$ and $q_2\bar{q}_4$.
Similarly, the tetraquark systems can be also rewritten in the $|(q_1\bar{q}_4)(q_2\bar{q}_3)\rangle$ bases.
We will not go into details here, and the transformation relationship between different bases can be obtained by  Eq.(\ref{13-12}).

Thus, the corresponding $4\times4$ matrix element of the chromomagnetic interaction for the scalar tetraquark systems with the quantum number $0^{+}$ is
\begin{widetext}
\begin{equation}\label{cm0}
H_{cm}=-\left(
\begin{array}{cccc}
16(v_{13}+v_{24})& 0 & 0
&
\left.
\begin{array}{c}
-\frac{4}{3}\sqrt{6}(v_{12}+v_{34}\\
+v_{14}+v_{23})
\end{array}
\right.\\[1.5em]
0 & -\frac{16}{3}(v_{13}+v_{24}) &\left.
\begin{array}{c}
-\frac{4}{3}\sqrt{6}(v_{12}+v_{34}\\
+v_{14}+v_{23})
\end{array}
\right. &
\left.
\begin{array}{c}
-\frac{8}{3}\sqrt{2}(v_{12}+v_{34}\\
-v_{14}-v_{23})
\end{array}
\right.\\[1.5em]
0 &
\left.
\begin{array}{c}
-\frac{4}{3}\sqrt{6}(v_{12}+v_{34}\\
+v_{14}+v_{23})
\end{array}
\right.
&-2(v_{13}+v_{24})&
\left.
\begin{array}{c}
\frac{4}{3}\sqrt{3}(v_{12}+v_{34})\\
-\frac{14}{3}\sqrt{3}(v_{14}+v_{23})
\end{array}
\right.\\[1.5em]
\left.
\begin{array}{c}
-\frac{4}{3}\sqrt{6}(v_{12}+v_{34}\\
+v_{14}+v_{23})
\end{array}
\right. &
\left.
\begin{array}{c}
-\frac{8}{3}\sqrt{2}(v_{12}+v_{34}\\
-v_{14}-v_{23})
\end{array}
\right. &
\left.
\begin{array}{c}
\frac{4}{3}\sqrt{3}(v_{12}+v_{34})\\
-\frac{14}{3}\sqrt{3}(v_{14}+v_{23})
\end{array}
\right. &
\left.
\begin{array}{c}
\frac{2}{3}(v_{13}+v_{24}+4v_{12}+4v_{34}\\
+14v_{14}+14v_{23})
\end{array}
\right.
\end{array}
\right).
\end{equation}
\end{widetext}
In a similar way, we can get the matrix element of $H_0$ for scalar tetraquark states which include both the chromoelectric interaction and the effective quark mass:
\begin{equation}\label{ce0}
H_0=-\left(
\begin{array}{cccc}
\cal{A}& 0 & \cal{C} & 0\\
0 & \cal{A} &0 & \cal{C}\\
\cal{C}& 0 & \cal{B} & 0\\
0 & \cal{C} &0 & \cal{B}\\
\end{array}
\right),
\end{equation}
with the notes,
\begin{eqnarray}
{\cal{A}} &\equiv& -(m_{13}+m_{24}),\nonumber\\
{\cal{B}} &\equiv&\frac{1}{8}(m_{13}+m_{24})-\frac{1}{4}(m_{12}+m_{34})-\frac{7}{8}(m_{14}+m_{23}),\nonumber\\
{\cal{C}} &\equiv& \frac{\sqrt{2}}{4}(m_{12}+m_{34}-m_{14}-m_{23}).
\label{ces3}
\end{eqnarray}
For the axial vector tetraquark systems with quantum number $J^P=1^+$, the basis vectors can be expressed as(labeled as $\beta$):
\begin{equation}\label{1324basis1}
\left.
\begin{array}{ll}
\hspace{-4mm}\beta_1\equiv |(q_1\bar{q}_3)^1_0\otimes(q_2\bar{q}_4)^1_1\rangle, & \beta_2\equiv|(q_1\bar{q}_3)^1_1\otimes(q_2\bar{q}_4)^1_0\rangle,\\[0.5em]
\hspace{-4mm}\beta_3\equiv|(q_1\bar{q}_3)^1_1\otimes(q_2\bar{q}_4)^1_1\rangle, & \beta_4\equiv|(q_1\bar{q}_3)^8_0\otimes(q_2\bar{q}_4)^8_1\rangle,\\[0.5em]
\hspace{-4mm}\beta_5\equiv|(q_1\bar{q}_3)^8_1\otimes(q_2\bar{q}_4)^8_0\rangle, & \beta_6\equiv|(q_1\bar{q}_3)^8_1\otimes(q_2\bar{q}_4)^8_1\rangle.\\
\end{array}
\right.
\end{equation}

For tetraquark systems with the quantum number $1^+$, above all basis states introduced have a definite charge conjugation if $q_1$ and $q_3$ (or $q_2$ and $q_4$) are the same flavor.
The basis $\beta_3$ and $\beta_6$ have positive charge conjugation and $\beta_1$, $\beta_2$, $\beta_4$, $\beta_5$ have negative charge conjugation.
Now, the chromomagnetic interaction Hamiltonian $H_{cm}$ acting on this basis vectors (\ref{1324basis1}) can be written the following blocks matrix,
\begin{equation}\label{cm1}
H_{cm}=-\left(
\begin{array}{ccc}
A_{11} & A_{12}\\
A_{21} & A_{22}
\end{array}
\right),
\end{equation}
with $3\times3$ dimensional sub-matrices,
\begin{widetext}
\begin{equation}
A_{11}=\left(
\begin{array}{ccc}
\frac{16}{3}(3v_{13}-v_{24})& 0 & 0\\[1.5em]
0 & -\frac{16}{3}(v_{13}-3v_{24}) & 0\\[1.5em]
0 & 0 & -\frac{16}{3}(v_{13}+v_{24})
\end{array}
\right),
\end{equation}

\begin{equation}
A_{22}= \left(
\begin{array}{ccc}
-2v_{13}+\frac{2}{3}v_{24} &
\left.
\begin{array}{c}
-\frac{4}{3}(v_{12}+v_{34})\\
+\frac{14}{3}(v_{14}+v_{23})
\end{array}
\right.&
\left.
\begin{array}{c}
\frac{4\sqrt{2}}{3}(v_{12}-v_{34})\\
+\frac{14\sqrt{2}}{3}(v_{14}-v_{23})
\end{array}
\right.\\[1.5em]
\left.
\begin{array}{c}
-\frac{4}{3}(v_{12}+v_{34})\\
+\frac{14}{3}(v_{14}+v_{23})
\end{array}
\right. &
\frac{2}{3}v_{13}-2v_{24} &
\left.
\begin{array}{c}
-\frac{4\sqrt{2}}{3}(v_{34}-v_{12})\\
+\frac{14\sqrt{2}}{3}(v_{14}-v_{23})
\end{array}
\right.\\[1.5em]
\left.
\begin{array}{c}
\frac{4\sqrt{2}}{3}(v_{12}-v_{34})\\
+\frac{14\sqrt{2}}{3}(v_{14}-v_{23})
\end{array}
\right. &
\left.
\begin{array}{c}
-\frac{4\sqrt{2}}{3}(v_{34}-v_{12})\\
+\frac{14\sqrt{2}}{3}(v_{14}-v_{23})
\end{array}
\right. &
\left.
\begin{array}{c}
\frac{2}{3}(v_{13}+v_{24})+\frac{4}{3}(v_{12}+v_{34})\\
+\frac{14}{3}(v_{14}+v_{23})
\end{array}
\right.
\end{array}
\right),
\end{equation}

\begin{equation}
A_{12}= A_{21}^{\dagger}=\left(
\begin{array}{ccc}
0 & \left.
\begin{array}{c}
\frac{4\sqrt{2}}{3}(v_{12}+v_{34}
+v_{14}+v_{23})
\end{array}
\right.& \left.
\begin{array}{c}
-\frac{8}{3}(v_{12}-v_{34}
-v_{14}+v_{23})
\end{array}
\right.\\[1.5em]
\left.
\begin{array}{c}
\frac{4\sqrt{2}}{3}(v_{12}+v_{34}
+v_{14}+v_{23})
\end{array}
\right. & 0 & \left.
\begin{array}{c}
-\frac{8}{3}(v_{34}-v_{12}
+v_{23}-v_{14})
\end{array}
\right.\\[1.5em]
\left.
\begin{array}{c}
-\frac{8}{3}(v_{12}-v_{34}
-v_{14}+v_{23})
\end{array}
\right. & \left.
\begin{array}{c}
-\frac{8}{3}(v_{34}-v_{12}
+v_{23}-v_{14})
\end{array}
\right. & \left.
\begin{array}{c}
-\frac{4\sqrt{2}}{3}(v_{12}+v_{34}
-v_{14}-v_{23})
\end{array}
\right.
\end{array}
\right).
\end{equation}
\end{widetext}
And the corresponding chromoelectric interaction related term can be given by
\begin{equation}\label{ce1}
H_{0}=-\left(
\begin{array}{cccccc}
\cal{A}& 0 &0 & \cal{C} & 0 &0\\
0 & \cal{A} &0 &0 & \cal{C} &0\\
0& 0 & \cal{A} & 0 &0 &\cal{C}\\
\cal{C}& 0 & 0 &\cal{B} & 0 &0\\
0 & \cal{C} & 0 &0 &\cal{B} &0\\
0 & 0 & \cal{C} & 0 &0& \cal{B}\\
\end{array}
\right),
\end{equation}
where these symbols ($\cal{A}$, $\cal{B}$, and $\cal{C}$) are consistent with the expressions in the previous Eq.(\ref{ces3}).

Similarly, for $J^{P}=2^{+}$ states, the basis vectors $\gamma$ of tetraquark systems are easier to get, as shown following,
\begin{equation}\label{1324basis2}
\left.
\begin{array}{cc}
\hspace{-2mm}\gamma_1\equiv|(q_1\bar{q}_3)^1_1\otimes(q_2\bar{q}_4)^1_1\rangle, &
\gamma_2\equiv|(q_1\bar{q}_3)^8_1\otimes(q_2\bar{q}_4)^8_1\rangle.\\
\end{array}
\right.
\end{equation}
The important thing is that under this set of basis vectors, the total spin of any quark antiquark pairs are always one.
Besides, $\gamma_1$ is similar to the molecular state formed by mesons and mesons, while $\gamma_2$ is more inclined to the compact tetraquark state.
In a similar way, the corresponding chromomagnetic interaction Hamiltonian $H_{cm}$ acting on this basis (\ref{1324basis2}) can be written by,
\begin{widetext}
\begin{equation}\label{cm2}
H_{cm}=-\left(
\begin{array}{ccc}
-\frac{16}{3}(v_{13}+v_{24})
&\left.
\begin{array}{c}
\frac{4}{3}\sqrt{2}(v_{12}+v_{34}
-v_{14}-v_{23})
\end{array}
\right.\\[1.5em]
\left.
\begin{array}{c}
\frac{4}{3}\sqrt{2}(v_{12}+v_{34}
-v_{14}-v_{23})
\end{array}
\right.
&\left.
\begin{array}{c}
\frac{2}{3}(v_{13}+v_{24})
-\frac{4}{3}(v_{12}+v_{34})-\frac{14}{3}(v_{14}+v_{23})
\end{array}
\right.
\end{array}
\right),
\end{equation}
\end{widetext}
and the corresponding $H_0$ composed of both the chromoelectric interaction term and the effective constituent quark mass is
\begin{equation}\label{ce2}
H_{0}=-\left(
\begin{array}{cccc}
\cal{A} & \cal{C}\\
\cal{C}& \cal{B}\\
\end{array}
\right).
\end{equation}
Still, $\cal{A}$, $\cal{B}$, and $\cal{C}$ are same with the expressions in Eq.(\ref{ces3}).

With these derivations, we can estimate the mass spectra of any tetraquark state by diagonalizing Eq.(\ref{m}).
%It is worth emphasizing that the coefficients $v_{ij}$ and $c_{ij}$ should be replaced by integrals containing the chosen form factor of the corresponding interaction and the orbital wave functions.
In this paper, we only consider the $S$-wave tetraquark systems and the specific calculation of the mass spectra will be carried out in the next section.

\section{Mass spectra and decay modes}

\subsection{the model parameters $m_{ij}$ and the coupling strengths $v_{ij}$ between quarks}

In order to estimate the mass spectrum splitting of the open-heavy tetraquark $qq\bar{q}\bar{Q}$ ($q=u,d,s$ and $Q=c,b$) systems, we need to determine the values of the relevant parameters.
Considering the breaking of flavor symmetry, we will use the new expression $n=u,d$ to further distinguish between $u$, $d$ and $s$ quarks.
Physically, the effective mass $m_i$ and the coupling strengths $v_{ij}$, $c_{ij}$ can be obtained by using the Hamiltonian Eq.(\ref{Ha}) to fit experimental masses of meson and baryon ground states.
However, different hadron systems have their own parameter values even in the case of the same phenomenal model.
Therefore, we can use a compromise method to combine the effective mass of the constituent quarks and the interaction of the chromoelectricity into one parameter.
The model parameters are only $m_{ij}$ and the chromomagnetic interaction coupling strengths $v_{ij}$ in Eq.(\ref{m}).

For a meson system composed of $q\bar{q}$, the mass spectrum can be obtained by Eq.(\ref{m}):
\begin{equation}\label{Mesonmass}
M_{M}=
\left\{
\begin{array}{ll}
m_{q\bar{q}}-16v_{q\bar{q}} & \text{for} \,\,\, \,S=0,\\[0.8em]
m_{q\bar{q}}+\frac{16}{3}v_{q\bar{q}} & \text{for} \,\,\, \,S=1,
\end{array}
\right.
\end{equation}
where the $M_{M}$ represents the mass of the meson, the $m_{q\bar{q}}$ and $v_{q\bar{q}}$ are model parameters.
Then, this mass splitting can be effectively calculated by the mass formula Eq.(\ref{Mesonmass}).
Now selecting the corresponding experimental values of scalar meson and vector meson mass~\cite{Zyla:2020zbs}, the model parameters can be extracted and displayed in Table~\ref{mesonpm} and Table~\ref{mesonpv}.
Since the spin-singlet state of $s\bar{s}$ and the spin-triplet state of $b\bar{c}$ are not yet known experimentally, the parameters $v_{s\bar{s}}$ and $v_{c\bar{b}}$ are obtained via similar strategy as shown in Ref.~\cite{Wu:2018xdi}.
In Table~\ref{mesonpm} and Table~\ref{mesonpv}, it is worth noting that we need to declare $v_{q_1\bar{q}_2}=v_{\bar{q}_1q_2}$ and $m_{q_1\bar{q}_2}= m_{\bar{q}_1q_2}$.
It is generally believed that the two configurations, $q_1\bar{q}_2$ and $\bar{q}_1q_2$, have the same flavor and spatial structure.

For $S$-wave baryons with the total spin $S=3/2$, the mass spectrum is given by
\begin{eqnarray}\label{Baryonmass}
M_{B}=\frac{1}{2}(m_{12}+m_{13}+m_{23})+\frac{8}{3}(v_{12}+v_{13}+v_{23}),
\end{eqnarray}
where the $M_{B}$ represents the mass of the baryon, the $m_{ij}$ and $v_{ij}$ are model parameters.
For baryons ($q_1q_2q_3$) with the total spin $S=1/2$, the basis can be constructed by $|(q_1q_2)_1(q_3)_{1/2}\rangle$ and $|(q_1q_2)_0(q_3)_{1/2}\rangle$.
It can be seen that this set of bases include symmetric or antisymmetric coupling of the first two quarks.
Therefore, the mass spectrum of the baryon of  the total spin $S=1/2$ can be written as
\begin{eqnarray}\label{Baryonmass1}\nonumber
M_{B}&=&\frac{1}{2}(m_{12}+m_{13}+m_{23})\\
&+&\frac{8}{3}
\left[
\begin{array}{cc}
v_{12}-2v_{13}-2v_{23}& \sqrt{3}(v_{23}-v_{13})\\
\sqrt{3}(v_{23}-v_{13}) & -3v_{12}
\end{array}
\right].
\end{eqnarray}
For the total spin $S = 1/2$ baryons with two identical quarks, the Eq.(\ref{Baryonmass1}) can be simplified by
\begin{equation}\label{Baryonmass2}
M_{B}=\frac{1}{2}(m_{12}+2m_{13})+\frac{8}{3}(v_{12}-4v_{13}).
\end{equation}
\begin{table}[!t]
\caption{\label{mesonpm} Parameters $m_{ij}$ (in MeV) for quark and antiquark system.}
\begin{ruledtabular}
\begin{tabular}{lcccc}
 & n & s & c & b\\
\hline
$\bar{n}$&616.34&&&\\
$\bar{s}$&792.17&$963.43$&&\\
$\bar{c}$&1975.11&2076.24&3068.67&\\
$\bar{b}$&5313.36&5403.28&$6327.40$&$9444.91$\\
\end{tabular}
\end{ruledtabular}
\end{table}

\begin{table}[t]
\caption{\label{mesonpv} Coupling strength parameters $v_{ij}$ (in MeV) for quark and antiquark system.
 }
\begin{ruledtabular}
\begin{tabular}{lcccc}
 & n & s & c & b\\
\hline
$\bar{n}$&29.798&&&\\
$\bar{s}$&18.656&$10.506$&&\\
$\bar{c}$&6.591&6.743&5.298&\\
$\bar{b}$&2.126&2.273&$3.281$&$2.888$\\
\end{tabular}
\end{ruledtabular}
\end{table}
Then, the model parameters ($m_{qq}$ and $v_{qq}$) corresponding to the baryon system composed of quark-quark ($qq$) can also be extracted by using the above approach.
Experimental values of baryon mass spectra are taken from the Particle Data Group (PDG) \cite{Zyla:2020zbs}.
The values of the extracted parameters $m_{qq}$ and $v_{qq}$ are displayed in Table~\ref{baryonpm} and Table~\ref{baryonpv}, respectively.
Here we still state $m_{qq}=m_{\bar{q}\bar{q}}$ and $m_{qq}=m_{\bar{q}\bar{q}}$ in Table~\ref{baryonpm} and Table~\ref{baryonpv}.
In addition, we can bring in these parameters to get the theoretical values of the mass spectra of mesons and baryons. The results are listed in Table~\ref{The-Exp}.
It shows good agreement with the the experimental data.

For the tetraquark systems, there are the known complication of spatial structures and different color-spin arrangements. We assume that the interaction potential is the summation of the two-body interactions.
In fact, three-body or many-body interactions are not so universal and important~\cite{PhysRevLett.122.103001}.
Therefore, the mass spectra and possible decay channels of the corresponding tetraquark states can be obtained through these extracted model parameters ($m_{ij}$ and $v_{ij}$) and the matrix elements of the tetraquark states solved previously.

\begin{table}[htbp]
\caption{\label{baryonpm} Parameters $m_{ij}$ (in MeV) for diquark system.}
\begin{ruledtabular}
\begin{tabular}{lcc}
& n & s \\
\hline
$n$&723.86&\\
$s$&904.83&$1080.59$\\
$c$&2085.58&2185.99\\
$b$&5413.07&5510.83\\
\end{tabular}
\end{ruledtabular}
\end{table}

\begin{table}[htbp]
\caption{\label{baryonpv} The values of  the coupling strength parameters $v_{ij}$ (in MeV) for diquark system.}
\begin{ruledtabular}
\begin{tabular}{lcc}
& n & s  \\
\hline
$n$&18.277&\\
$s$&12.824&$6.445$\\
$c$&4.063&4.148 \\
$b$&1.235&1.304\\
\end{tabular}
\end{ruledtabular}
\end{table}

\begin{table}[htbp]
\caption{\label{The-Exp}The Mass spectra (in MeV) of ground state mesons and baryons. And the corresponding experimental values come from the Particle Data Group~\cite{Zyla:2020zbs}.}
\vspace{0.2cm}
\centering
\begin{tabular}{ccc|ccc}
\hline\hline
Hadron &Expt. & Theo. & Hadron & Expt. & Theo.  \\
\hline\hline
$\pi^+$ &$139.57$  & $139.57$ & $\rho^+$ & $775.26$  & $775.26$  \\
$K$ & $493.68$  & $493.68$  & $K^\ast$ & $891.66$ & $891.67$\\
$-$ &   &   & $\phi$ & $1019.46$ & $1019.46$\\
$D^-$ & $1869.65$ & $1869.65$ & $D^{\ast -}$ & $2010.26$ & $2010.26$\\
$D_s^-$ & $1968.34$ & $1968.34$ & $D_s^{\ast -}$ & $2112.2$ & $2112.2$\\
$\eta_c$ & $2983.90$ & $2983.90$ & $J/\Psi$ & $3096.92$  & $3096.93$\\
$B^+$ & $5279.34$ & $5279.34$ & $B^\ast$ & $5324.7$ & $5324.7$\\
$B_s^0$ & $5366.90$ & $5366.91$ & $B_s^\ast$ & $5415.4$ & $5415.4$\\
$B_c^+$ & $6274.9$ & $6274.9$ & $-$ & \\
$\eta_{b}$ & $9398.7$ & $9398.7$ & $\Upsilon$ & $9460.3$ & $9460.3$\\
%\hline
$N$ & $939.57$ & $ 939.57$ & $\Delta$ & $1232$ & $1232.01$\\
$\Sigma^+$ & $1189.37$ &1178.71 &$\Sigma^{\ast +}$ & $1382.8$ & $1383.89$\\
$\Xi^0$ & $1314.86$ & $1325.52$ &$\Xi^{\ast 0}$ & $1531.80$ & $1530.71$  \\
$-$ & &  &$\Omega^{-}$ & $1672.45$  & 1672.45  \\
$\Sigma_c^+$ & $2452.9$ & 2452.9 & $\Sigma_c^{\ast ++}$ & $2517.9$ & 2517.9  \\
$\Xi^\prime_c$ & $2579.2$ &2578.61  &$\Xi^{\ast +}_c$ & $2645.56$ & 2644.29  \\
$\Omega_c^0$ & $2695.2$ & 2699.23 &$\Omega_c^{\ast 0}$ & $2765.9$ &2765.59  \\
%$\Xi_{cc}^{++}$ & $3621.2$ & 3621.2 &$\Xi_{cc}^{\ast ++}$ & $3754/3706^\bullet$ &  \\
%$\Omega_{cc}^+$ & $3832^\bullet/3815$ &  &$\Omega_{cc}^{\ast +}$ & $3883^\bullet/3876^\bullet$ &  \\
$\Sigma_{b}^+$ & $5810.56$ & 5810.57 &$\Sigma_{b}^{\ast +}$ & $5830.32$ & 5830.33 \\
$\Xi_{b}^\prime$ & $5935.02$ & 5935.02 & $\Xi_{b}^{\ast}$ & $5955.33$ & 5955.33\\
$\Omega_{b}^-$ & $6046.1$ & $6054.4$  &$-$ & &      \\
%$\Xi_{bb}^{++}$ & $10314/10340 ^\bullet$ & &$\Xi_{bb}^{++}$ & $10339/10367 ^\bullet$ &  \\
%$\Omega_{bb}^+$ & $10447^\bullet/10454$ & & $\Omega_{bb}^{\ast +}$ & $10467^\bullet/10484^\bullet$  &  \\
\hline\hline
\end{tabular}
\end{table}

\subsection{Tetraquark states with $nn\bar{n}\bar{c}$,\, $ss\bar{s}\bar{c}$\, $nn\bar{s}\bar{c}$,\, $ss\bar{n}\bar{c}$,\, $sn\bar{n}\bar{c}$,\, $sn\bar{s}\bar{c}$ configuration}

%---------------------------------------------------------------------------------------------------------------------
\begin{table*}[!htbp]%[!htbp]
\caption{\label{T1}The mass $M$ (in MeV) and the amplitudes of the basis vectors for the tetraquark $nn\bar{n}\bar{c}$, $ss\bar{s}\bar{c}$, $nn\bar{s}\bar{c}$, $ss\bar{n}\bar{c}$, $sn\bar{n}\bar{c}$ and $sn\bar{s}\bar{c}$ systems with the different quantum numbers $J^P=0^+$, $1^+$ and $2^+$.
The Hamiltonian and the basis have been introduced in Sec.II.
The amplitudes $c_i$ in the (a), (b), (c), (d), (e), (f) parts correspond to the $|(q_1\bar{q}_3)(q_2\bar{q}_4)\rangle$ configure, while the amplitudes $c_i$ in the (e$^\prime$) and (f$^\prime$) parts correspond to the $|(q_1\bar{q}_4)(q_2\bar{q}_3)\rangle$ configure.}
\vspace{0.2cm}
\centering
\begin{tabular}{cccc|cccc}
 \hline \hline
System & $J^{P}$ & $M$(MeV) & $c_i$ &System & $J^{P}$ & $M$(MeV) & $c_i$     \\
 \hline\hline
$nn\bar{n}\bar{c}$
&$0^{+}$ & 2953.9& (0.04, -0.72, -0.69, -0.09) &        $ss\bar{s}\bar{c}$ &$0^{+}$ & 3229.4& (0.13, -0.67, -0.73, -0.01)\\
(a)&        & 2707.2& (-0.07, 0.63, -0.59, -0.49) &   (b) &                      & 3139.6& (-0.01, 0.62, -0.57, -0.54) \\
&        & 2238.3& (-0.64, -0.27, 0.33, -0.64) &    &                      & 2885.3& (-0.63, -0.36, 0.23, -0.65) \\
&        & 1836.5& (-0.76, 0.13, -0.26, 0.58) &     &                      & 2667.6& (-0.76, 0.18, -0.30, 0.54) \\
&$1^{+}$ & 2885.7 & ({  -0.04, -0.23, 0.66, 0.67, -0.25, -0.06})&  &$1^{+}$& 3186.1& ({-0.09, 0.03, 0.64, 0.54, -0.45, -0.30})\\
&        & 2764.1 &({  0.05, -0.19, -0.70, 0.52, -0.36, 0.27})&    &       & 3169.8& (0.03, -0.09, 0.63, -0.49, 0.51, -0.32)\\
&        & 2744.4 &({ 0.11, 0.62, 0.04, -0.11, -0.72, -0.26})&     &       & 3140.0& (-0.31, -0.49, -0.03, 0.56, 0.57, 0.12)\\
&        & 2575.1 &({  0.01, -0.68, -0.09, -0.32, -0.33, -0.57})&  &       & 2954.6& (0.27, -0.70, -0.27, -0.10, -0.25, -0.54)\\
&        & 2344.5 &({  0.59, -0.24, 0.23, -0.33, -0.27, 0.60})&    &       & 2947.7& (0.52, -0.39, 0.34, -0.04, -0.14, 0.66) \\
&        & 2028.4 &({  0.80, 0.10, -0.10, 0.27, 0.31, -0.42})&     &       & 2835.0& (-0.74, -0.33, 0.10, -0.39, -0.36, 0.24)\\
&$2^{+}$ & 2857.1         & (-0.58, 0.82)  &                       &$2^{+}$& 3227.4& (-0.58, 0.82)\\
&        & 2749.8        & (-0.82, -0.58)  &                       &       & 3083.8& (-0.82, -0.58) \\
\hline
$nn\bar{s}\bar{c}$
&$0^{+}$ & 3041.6& (-0.03, 0.72, 0.69, 0.10) &        $ss\bar{n}\bar{c}$ &$0^{+}$ & 3124.5& (0.11, -0.68, -0.72, -0.02)\\
(c)&        & 2799.9& (-0.10, 0.64, -0.60, -0.47) &   (d) &                    & 3037.2& (-0.01, 0.62, -0.57, -0.54) \\
&        & 2543.5& (0.65, 0.26, -0.34, 0.64) &    &                      & 2621.3& (-0.64, -0.35, 0.24, -0.65) \\
&        & 2224.4& (-0.76, 0.11, -0.23, 0.60)  &    &                    & 2337.4& (-0.76, 0.18, -0.30, 0.54) \\
&$1^{+}$ & 2975.3 & ({0.02, 0.20, -0.68, -0.64, 0.30, 0.06})&  &$1^{+}$  & 3078.0& ({-0.09, -0.002, 0.64, 0.57, -0.43, -0.28})\\
&        & 2866.3 &({-0.03, 0.36, 0.68, -0.53, 0.22, -0.26})&    &       & 3062.1& (0.02, -0.12, 0.63, -0.51, 0.47, -0.33)\\
&        & 2856.0 &({0.19, 0.52, -0.13, -0.09, -0.81, -0.14})&   &       & 3014.0& (-0.20, -0.57, 0.02, 0.50, 0.57, 0.25)\\
&        & 2687.2 &({0.04, -0.69, -0.03, -0.37, -0.30, -0.55})&  &       & 2816.6& (-0.09, 0.72, 0.29, 0.15, 0.32, 0.51)\\
&        & 2642.1 &({0.56, -0.28, 0.22, -0.32, -0.16, 0.66})&    &       & 2699.8& (0.60, -0.33, 0.30, -0.15, -0.25, 0.60) \\
&        & 2417.9 &({-0.81, -0.11, 0.08, -0.25, -0.31, 0.41})&   &       & 2512.2& (-0.77, -0.19, 0.14, -0.35, -0.32, 0.36)\\
&$2^{+}$ & 2965.7         & (-0.58, 0.82)  &                     &$2^{+}$& 3113.0& (-0.58, 0.82)\\
&        & 2871.1        & (-0.82, -0.58)  &                     &       & 2949.3& (-0.82, -0.58) \\
%\hline
%$ns\bar{n}\bar{c}$
%&$0^{+}$ & 3042.1& (0.05, -0.62, -0.78, -0.10) &        $ns\bar{s}\bar{c}$ &$0^{+}$ & 3138.5& (0.05, -0.61, -0.79, -0.11)\\
%&        & 2866.4& (-0.04, 0.71, -0.50, -0.49) &    &                    & 2962.9& (-0.05, 0.73, -0.50, -0.47) \\
%&        & 2475.3& (0.37, 0.33, -0.34, 0.80) &    &                  & 2751.9& (-0.37, -0.32, 0.34, -0.80) \\
%&        & 2044.5& (-0.93, 0.07, -0.16, 0.34)  &   &         & 2412.4& (-0.93, 0.06, -0.15, 0.34) \\
%&$1^{+}$ & 2982.6 & ({-0.05, -0.13, 0.49, 0.76, -0.38, -0.08})&  &$1^{+}$& 3082.3& (-0.03, -0.11, 0.46, 0.75, -0.44, -0.10)\\
%&        & 2904.2 &({-0.01, 0.06, 0.79, -0.42, 0.26, -0.36})&    &       & 3007.1& (0.08, 0.34, 0.67, -0.53, -0.23, -0.32)\\
%&        & 2882.6 &({-0.11, -0.56, 0.04, 0.27, 0.74, 0.23})&     &       & 3004.7& (0.15, 0.33, -0.47, -0.10, -0.79, 0.13)\\
%&        & 2697.3 &({0.02, -0.72, -0.21, -0.20, -0.29, -0.56})&  &       & 2851.5& (0.17, -0.14, 0.35, -0.14, -0.07, 0.90)\\
%&        & 2583.6 &({0.28, -0.37, 0.28, -0.33, -0.35, 0.69})&    &       & 2814.5& (0.14, -0.86, -0.03, -0.31, -0.31, -0.22) \\
%&        & 2213.8 &({-0.95, -0.05, 0.04, -0.16, -0.17, 0.17})&   &       & 2580.4& (-0.96, -0.07, 0.02, -0.16, -0.19, 0.13)\\
%&$2^{+}$ & 2988.0         & (-0.53, 0.85)  &                     &$2^{+}$& 3100.6& (-0.49, 0.88)\\
%&        & 2847.8        & (-0.85, -0.53)  &                     &       & 2974.2& (-0.88, -0.49) \\
\hline
$sn\bar{n}\bar{c}$
&$0^{+}$ & 3042.1& (0.10, -0.77, -0.63, -0.01) &        $sn\bar{s}\bar{c}$ &$0^{+}$ & 3138.5& (0.10, -0.78, -0.62, -0.005)\\
(e)&        & 2866.4& (-0.05, 0.53, -0.66, -0.53) &   (f) &                    & 2962.9& (-0.07, 0.52, -0.67, -0.52) \\
&        & 2475.3& (-0.85, -0.26, 0.20, -0.43) &    &                    & 2751.9& (-0.85, -0.27, 0.20, -0.42) \\
&        & 2044.5& (-0.52, 0.23, -0.37, 0.73)  &   &                     & 2412.4& (-0.52, 0.21, -0.36, 0.74) \\
&$1^{+}$ & 2982.5 & (-0.11, -0.18, 0.79, 0.51, -0.19, -0.20)&    &$1^{+}$& 3082.9 & (-0.12, -0.12, 0.82, 0.46, -0.20, -0.22)\\
&        & 2904.7 &(-0.06, -0.15, 0.44, -0.55, 0.67, -0.16)&    &        & 3009.6& (0.25, 0.42, -0.20, 0.18, -0.83, -0.02)\\
&        & 2884.2 &(-0.16, -0.63, -0.25, 0.44, 0.45, 0.35)&     &        & 3001.5& (-0.19, -0.46, -0.42, 0.71, -0.04, 0.28)\\
&        & 2695.0 &(-0.11, 0.69, 0.18, 0.27, 0.37, 0.52)&  &             & 2845.9& (0.74, -0.62, 0.10, -0.18, -0.15, 0.07)\\
&        & 2569.5 &(0.84, -0.18, 0.22, -0.09, -0.08, 0.44)&    &         & 2805.3& (0.35, 0.40, 0.24, 0.22, 0.28, 0.73) \\
&        & 2228.2 &(-0.49, -0.20, 0.20, -0.41, -0.41, 0.59)&   &         & 2595.2& (-0.47, -0.24, 0.19, -0.42, -0.41, 0.58)\\
&$2^{+}$ & 2988.0         & (-0.62, 0.78)  &                     &$2^{+}$& 3100.6& (-0.66, 0.75)\\
&        & 2847.8        & (-0.78, -0.62)  &                     &       & 2974.2& (-0.75, -0.66) \\
\hline
$sn\bar{n}\bar{c}$
&$0^{+}$ & 3042.1& (-0.05, 0.61, 0.78, 0.10) &        $sn\bar{s}\bar{c}$ &$0^{+}$ & 3138.5& (-0.05, 0.61, 0.79, 0.11)\\
(e$^\prime$)&        & 2866.4& (-0.04, 0.71, -0.50, -0.49) & (f$^\prime$)   &     & 2962.9& (-0.05, 0.72, -0.50, -0.47) \\
&        & 2475.3& (0.37, 0.33, -0.34, 0.80) &    &                  & 2751.9& (0.37, 0.32, -0.34, 0.80) \\
&        & 2044.5& (-0.93, 0.07, -0.16, 0.33)  &   &         & 2412.4& (-0.93, 0.06, -0.14, 0.34) \\
&$1^{+}$ & 2982.5 & (0.16, 0.05, 0.48, 0.38, -0.76, -0.12)&  &$1^{+}$& 3082.9& (0.14, 0.03, 0.44, 0.44, -0.75, -0.15)\\
&        & 2904.7 &(0.02, 0.02, -0.80, 0.21, 0.44, 0.35)&    &       & 3009.6& (-0.25, -0.14, 0.63, 0.29, 0.56, -0.35)\\
&        & 2884.2 &(0.46, 0.11, 0.10, -0.77, -0.27, 0.32)&     &     & 3001.5& (0.30, 0.12, 0.56, -0.76, -0.07, 0.001)\\
&        & 2695.0 &(0.79, 0.004, -0.25, 0.17, 0.17, -0.51)&  &       & 2845.9& (-0.06, -0.20, 0.30, 0.16, 0.07, 0.91)\\
&        & 2569.5 &(0.37, -0.32, 0.24, 0.38, 0.30, 0.68)&    &       & 2805.3& (0.91, -0.18, -0.06, 0.26, 0.28, -0.02) \\
&        & 2228.2 &(-0.06, -0.94, -0.07, -0.19, -0.18, -0.19)&   &   & 2595.2& (-0.07, -0.94, -0.06, -0.21, -0.18, -0.15)\\
&$2^{+}$ & 2988.0         & (-0.53, -0.85)  &                     &$2^{+}$& 3100.6& (-0.49, -0.88)\\
&        & 2847.8        & (0.85, -0.53)  &                     &       & 2974.2& (0.88, -0.49) \\
 \hline \hline
\end{tabular}
\end{table*}
%---------------------------------------------------------------------------------------------------------------------

%-----------------------------------------------------
\begin{figure*}[htbp]
\centering
\begin{minipage}{9.26cm}
    \includegraphics[width=0.9\textwidth,height=0.9\textwidth]{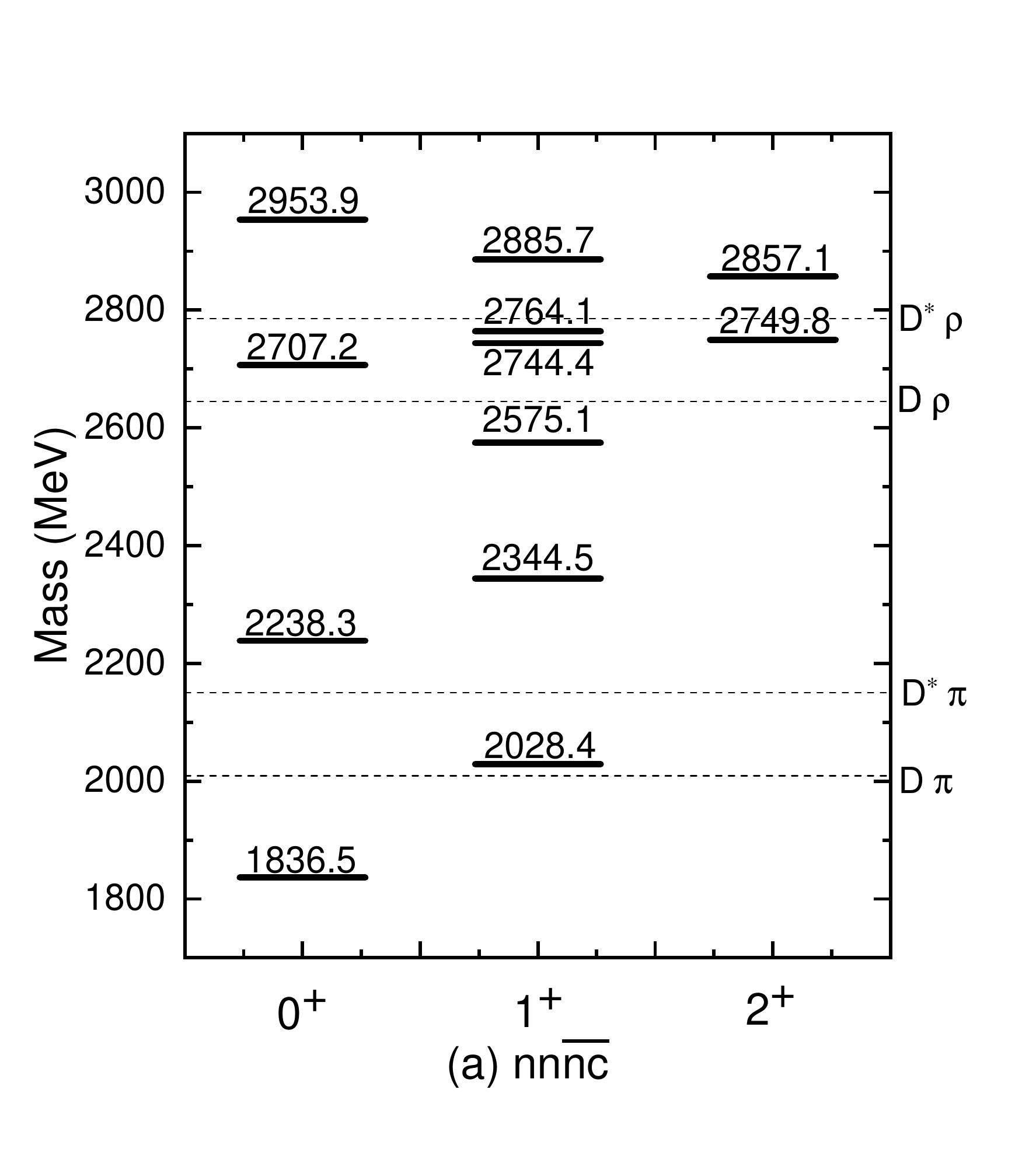}
  %\caption{}
  %\label{f2.12}
\end{minipage}
\hspace{-0.8cm}
\begin{minipage}{9.26cm}
    \includegraphics[width=0.9\textwidth,height=0.9\textwidth]{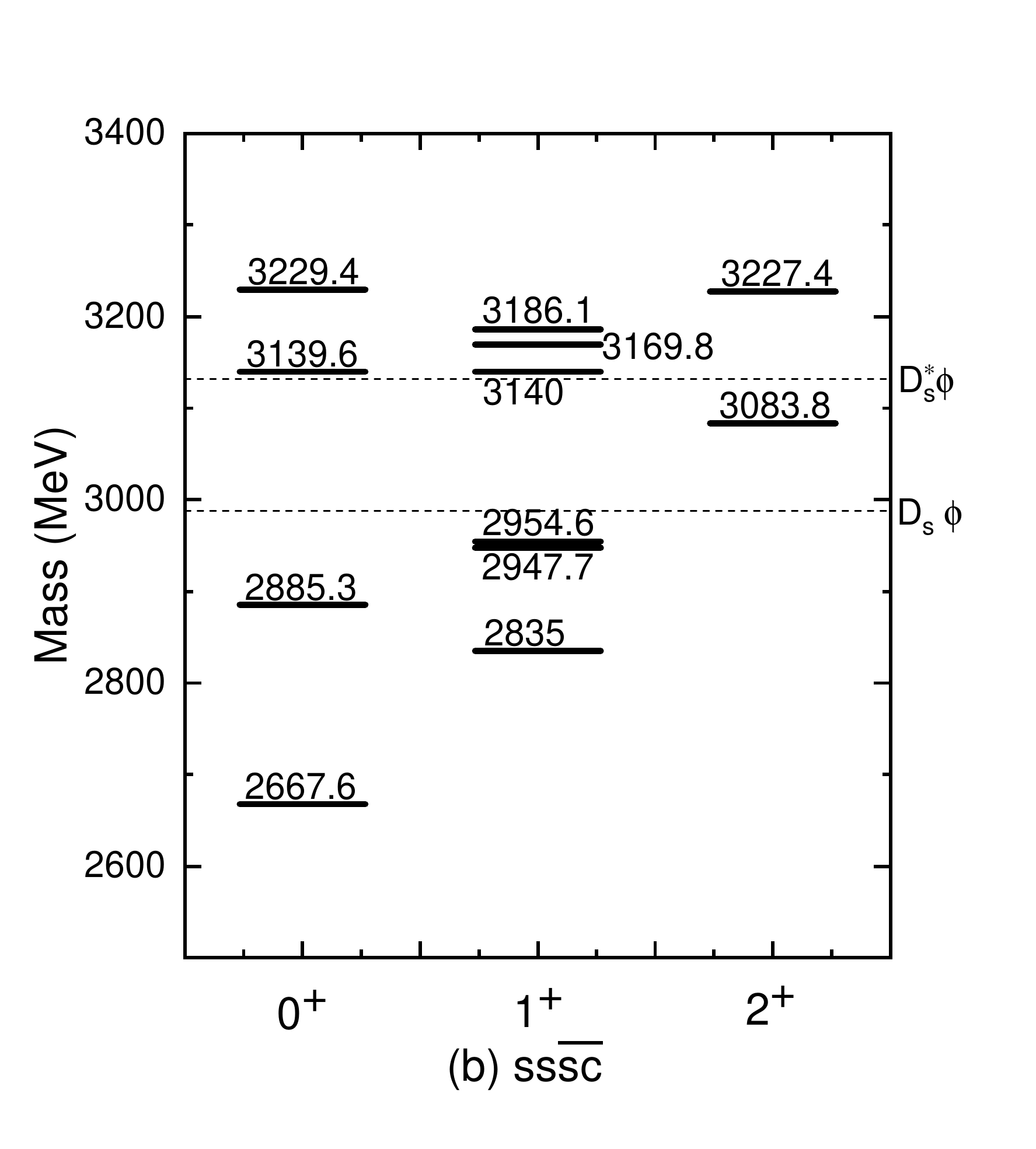}
\end{minipage}

\vspace{-1.2cm}

\begin{minipage}{9.26cm}
    \includegraphics[width=0.9\textwidth,height=0.9\textwidth]{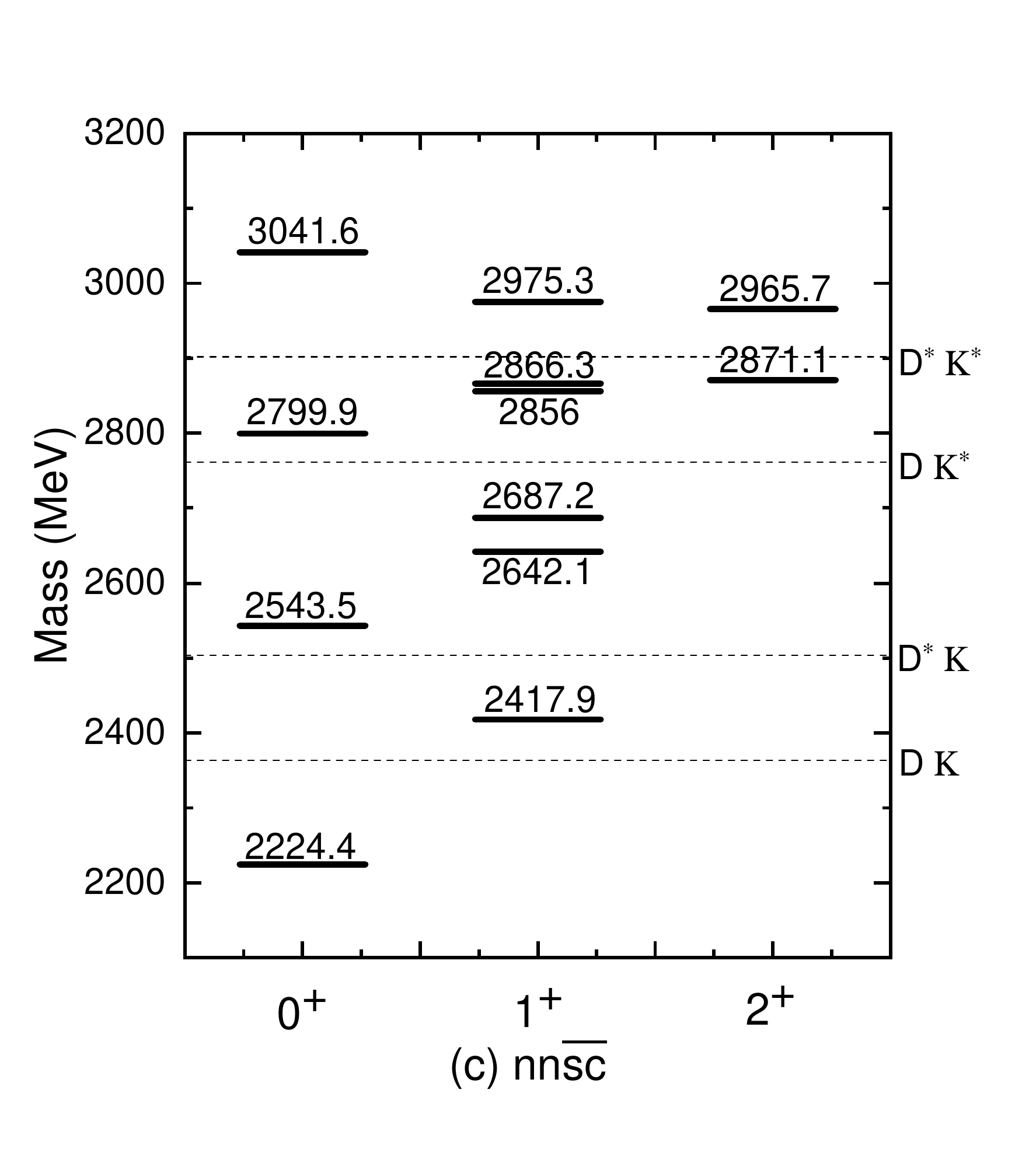}
\end{minipage}
\hspace{-0.8cm}
\begin{minipage}{9.26cm}
    \includegraphics[width=0.9\textwidth,height=0.9\textwidth]{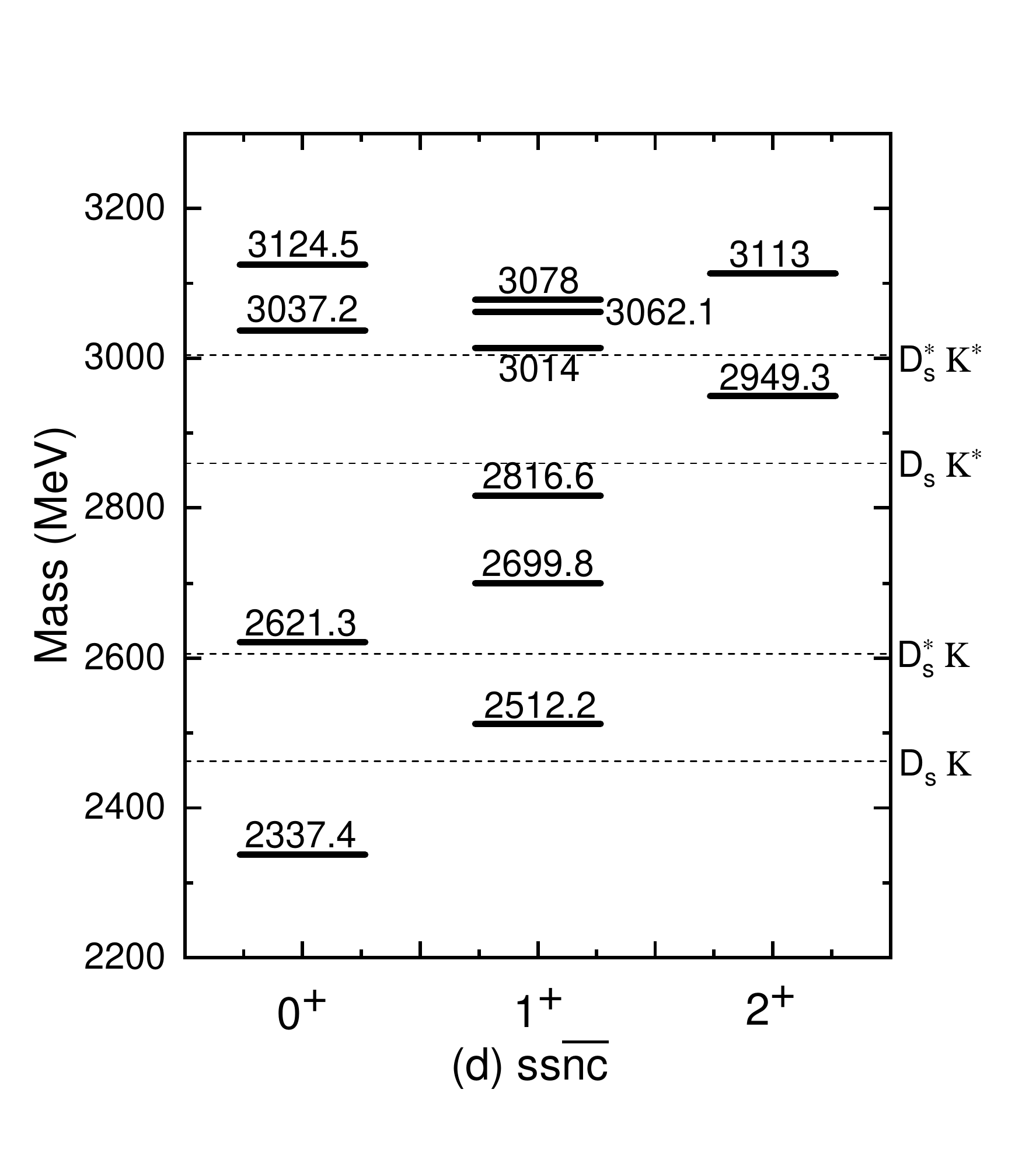}
\end{minipage}

\vspace{-1.2cm}

\begin{minipage}{9.26cm}
    \includegraphics[width=0.9\textwidth,height=0.9\textwidth]{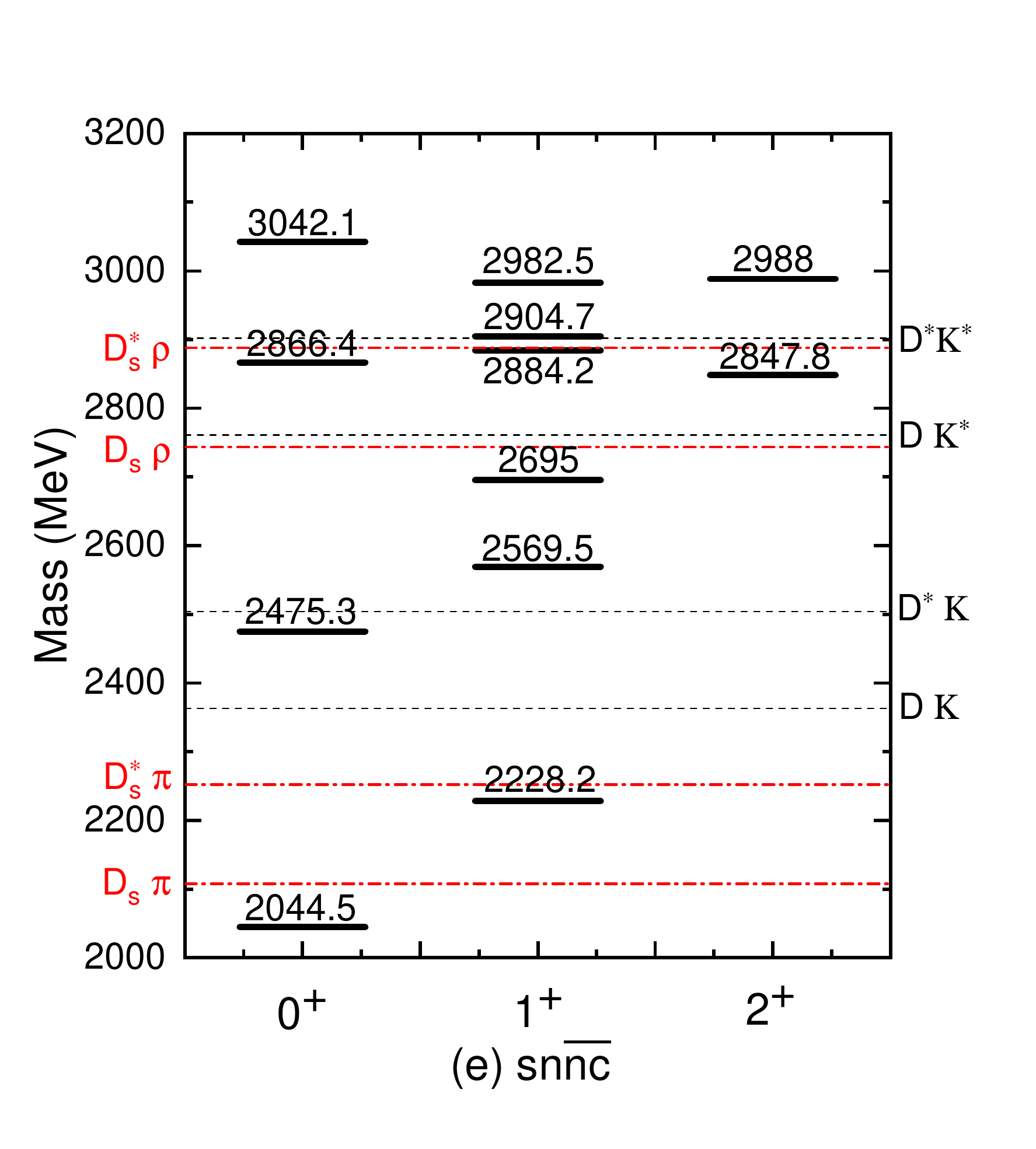}
\end{minipage}
\hspace{-0.8cm}
\begin{minipage}{9.26cm}
    \includegraphics[width=0.9\textwidth,height=0.9\textwidth]{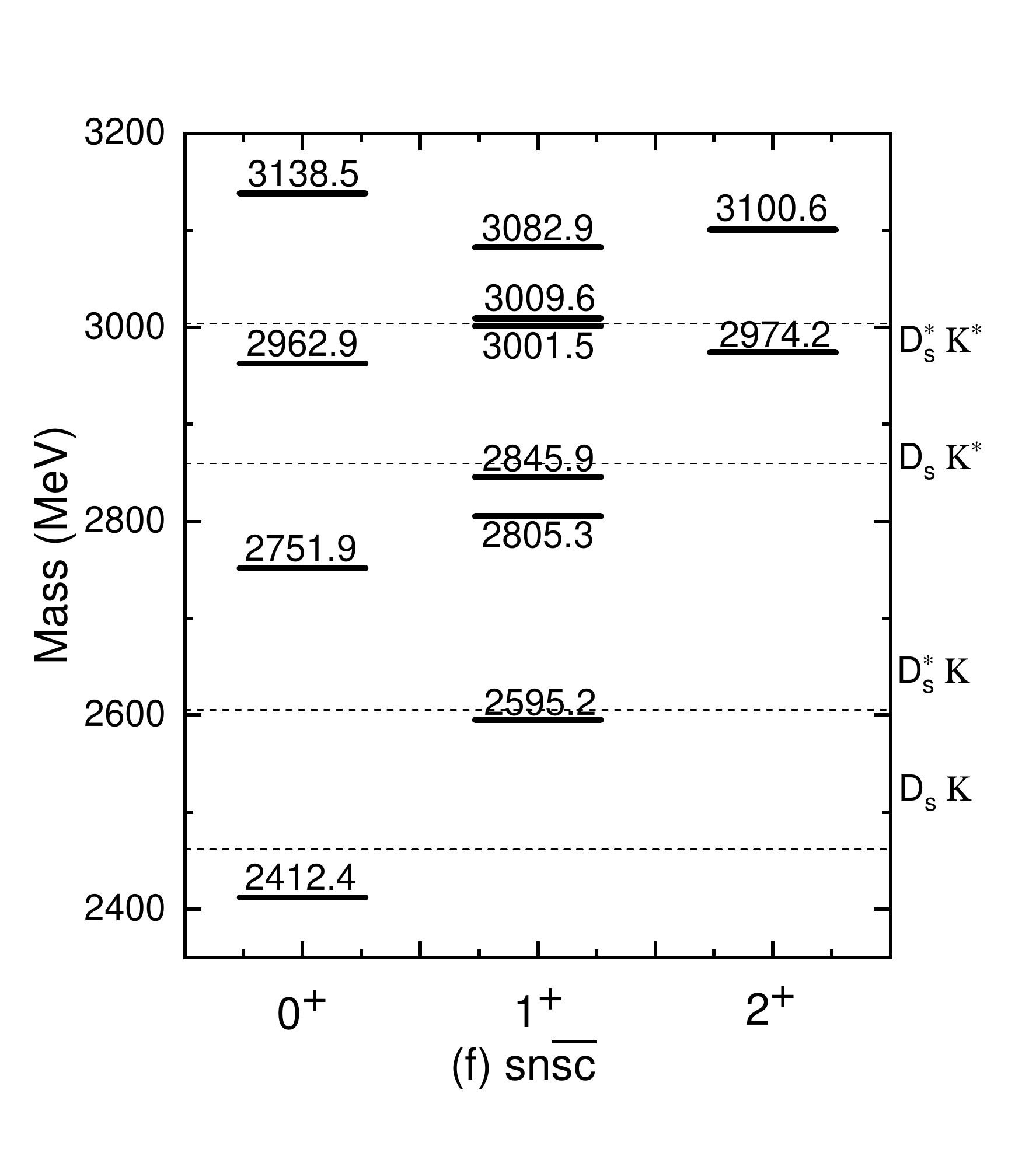}
\end{minipage}
\vspace{-0.5cm}
\caption{\label{f1}
The mass spectrum of the tetraquark system, $nn\bar{n}\bar{c}$, $ss\bar{s}\bar{c}$, $nn\bar{s}\bar{c}$, $ss\bar{n}\bar{c}$, $sn\bar{n}\bar{c}$ and $sn\bar{s}\bar{c}$.
The solid lines represent the mass spectrum of the possible tetraquark states.
The black dashed lines and the red dot-dashed lines are the thresholds of the possible meson-meson mass thresholds.
}
\end{figure*}
%-----------------------------------------------------

In this subsection, the low-lying $S$-wave states of the open charmed tetraquark systems are systematically investigated.
Now we present our calculation method for the masses which can be used to estimate the mass splitting among the tetraquark states.
Specifically, the mass spectrum can be obtained by solving the expectation value of Eq.(\ref{Ha1}).
The physical parameters and the tetraquark matrix elements have also been obtained in the previous sections. The quantum numbers of the systems may be $J^{P}$ = $0^+$, $1^{+}$, or $2^{+}$ for the $S$-wave tetraquark states.
Furthermore, the mass spectra of the open charmed tetraquark states can be easily calculated by diagonalizing the matrix Eq.(\ref{m}).
At the same time, we get the amplitudes of different color-spin wave functions, which can provide important information about the decay properties.
For convenience, the mass spectra and the amplitudes are collected in Table~\ref{T1}.
In addition, for easy comparison, the mass spectra of the open charm tetraquark systems are exhibited in Fig.~\ref{f1}.
Similarly, the corresponding meson-meson thresholds are also drawn in the figure.
The decay width depends on not only the branching fraction which is proportional to the square of the amplitude in the eigenvectors but also the phase space. For two-body decay, the partial decay width can be expressed as~\cite{Weng:2019ynv,Weng:2020jao},
\begin{equation}\label{dw}
\Gamma_i=\gamma_i \alpha \frac{k^{2L+1}}{M^{2L}}\cdot|c_i|^2,
\end{equation}
where $\gamma_i$ is a quantity determined by the decay dynamics and determined by the spatial wave functions of both initial and final states which are different for each decay process. $M$, $k$ and $c_i$ are mass of the parent particle, the momentum of the daughter particles in the rest frame of the parent particle, and the amplitude in the color-spin wave function basis, respectively.
The $\alpha$ is an effective coupling constant, and the parameter in the same structure of tetraquark state decay process can be approximately regarded as the same.

For the $nn\bar{n}\bar{c}$ system, the calculated mass spectra and the corresponding meson-meson thresholds can be seen in Fig.~\ref{f1}(a).
There are four possible $S$-wave tetraquark states with the quantum number $0^+$, and the mass spectrum of the lowest state is $1836.5$ MeV.
As can be seen from Table~\ref{T1}(a), the lowest state has a larger amplitude (-0.76) in the basis vector $\alpha_1$.
This implies that this state may be the strong coupling of the $D+\pi$ components.
The lowest state is obviously lower than the $D\pi$ threshold, however, which means that it be difficult to decay naturally to  $D+\pi$ mesons.
The smaller component (0.13) of the corresponding basis vector $\alpha_2$ implies a weak coupling to the $D^\ast\rho$ channel.
The basis vectors $\alpha_3$ and $\alpha_4$ correspond to the hidden color channels which generally do not participate in the strong interactional decay process.
And the hidden color part should decrease asymptotically when the tetraquark state is decomposed into two mesons.
The second state with the quantum number $0^+$ which mass is 2238.3MeV, above the $D\pi$ threshold.
It can naturally decay into $D+\pi$ mesons.
The larger amplitude (-0.64) of the basis vector $\alpha_1$  makes this state more inclined to
being dominant by the $D+\pi$ components.
The third and fourth states, 2707.2MeV and 2953.9MeV, have larger amplitudes of the basis vector $\alpha_2$.
Therefore, it can be considered that the main components of these two states are the strong coupling of the $D^\ast+\rho$ mesons.
Besides, the small amplitude of the basis vector $\alpha_1$  implies a negligible coupling strength to the $D\pi$ channel.
Since their mass is much greater than the $D\pi$ threshold, this also shows that they can naturally decay into $D+\pi$ mesons.

For the $nn\bar{n}\bar{c}$ system with the quantum number $1^+$, there are six possible tetraquark states.
What can be seen from Table~\ref{T1}(a), on the one hand, the two states, with the lowest mass 2028.4MeV and 2344.5MeV, have larger amplitudes of the basis vector $\beta_1$.
It implies that there is a strong coupling of $D^\ast+\pi$ mesons in these two states.
And the 2344.5MeV has enough mass to naturally decay into $D^\ast+\pi$ mesons.
On the other hand, the smaller amplitudes of the basis vectors $\beta_2$ and $\beta_3$ indicate that the two lowest states have decay channels of $D+\rho$ and $D^\ast +\rho$.
However, since these two tetraquark states have significantly smaller masses than the $D\rho$ or $D^\ast \rho$ thresholds, it may be difficult to complete such a decay mode.
The basis vectors $\beta_4$, $\beta_5$, and 
$\beta_6$ are the hidden color channels that directly do not participate in the decay of strong interactions.
Similarly, we can analyze the remaining four possible tetraquark states with the quantum number $1^+$.
The two states, 2575.1MeV and 2744.4MeV, have larger amplitudes of the basis vector $\beta_2$.
This implies that they have a main $D\rho$ strong coupling channel.
And 2744.4MeV is higher than the $D\rho$ threshold, so this state can naturally decay into $D\rho$ and $D^\ast\pi$ mesons.
For remaining states 2764.1MeV and 2885.7MeV of the quantum number $1^+$, from Table~\ref{T1} and Fig.~\ref{f1}(a), their main strong coupling channel is $D^\ast+\rho$, and 2885.7MeV is easier to decay to $D^\ast+\rho$ mesons because the mass is bigger than $D^\ast\rho$ threshold.
At the same time, these two states also include $D^\ast\pi$ and $D\rho$ decay channels.

In Fig.~\ref{f1}(a), there are two $S$-wave $nn\bar{n}\bar{c}$ states with a quantum number $2^+$, and their masses are close to the $D^\ast\rho$ threshold.
In addition, it can be found from Table~\ref{T1} that these two states have a nontrivial amplitude of basis vector $\gamma_1$, indicating that they can decay to the $D^\ast+\rho$ channel.
The basis vector $\gamma_2$ shows the hidden-color content which should decrease asymptotically when the tetraquark state decay into two mesons.
Comparing the masses of these two states with the threshold value of $D^\ast\rho$, it can be found that the state, 2857.1MeV, may decay substantially into  $D^\ast+\rho$ mesons, but the 2749.8MeV should be forbidden.

The structure of the $ss\bar{s}\bar{c}$ system is similar to that of the $nn\bar{n}\bar{c}$ system, but the related physical parameters are different.
The calculated results are exhibited in Table~\ref{T1}(b) and Fig.~\ref{f1}(b).
Since the $\eta$ meson is an isosinglet meson made of a mixture of up, down and strange quarks and their antiquarks, we did not draw the corresponding meson-meson threshold in Fig.~\ref{f1}(b).
But this does not affect our analysis of the possible decay channels.
This $ss\bar{s}\bar{c}$ system with the quantum number $0^+$ has four possible states, and the masses of the two lowest states are 2667.6MeV and 2885.3MeV.
The larger amplitudes of the basis vectors $\alpha_1$ allow them to decay into pseudoscalar-pseudoscalar channel, i.e., $D_s\eta$ channel.
Similarly what can be seen from Fig.~\ref{f1}(b), the two states, 3229.4MeV and 3139.6MeV, may decay substantially into $D_s^\ast+\phi$ mesons.
For the $J^P=1^+$ states, these two states, 2835MeV and 2947.7MeV, have a relatively large amplitude of the basis vectors $\beta_1$, corresponding to the possible decay of $D_s^\ast\eta$ channel.
The two states, 2954.6MeV and 3140MeV, may be the energy level splitting of the combination of $D_s+\phi$ mesons.
And the remaining two highest energy states can decay to $D_s^\ast+\phi$ mesons because they have a larger amplitude of the basis $\beta_3$ and a larger mass above the $D_s^\ast\phi$ threshold.
For the $S$-wave $ss\bar{s}\bar{c}$ states with the quantum number $2^+$, there are only two possible states, and their masses are respectively above and below $D_s^\ast\phi$ threshold.
Considering in conjunction with Table~\ref{T1}(b), it is easy to understand that these states can decay to $D_s^\ast+\phi$ mesons.
Moreover, it can be understood that the state with larger mass, 3227.4MeV, is easier to decay naturally into the $D^\ast\phi$ channel.
A state with a smaller mass, 3083.8MeV, is less likely to pass this decay.

Considering the $nn\bar{s}\bar{c}$ system, the calculated results are exhibited in Table~\ref{T1}(c) and Fig.~\ref{f1}(c).
For the quantum number $J^P=0^+$, there are four possible tetraquark states, and their masses are 2224.4MeV, 2543.5MeV, 2799.7MeV as well as 3041.6MeV in order.
The two states with lower masses (2224.4MeV and 2543.5MeV) have larger amplitudes of the basis vector $\alpha_1$, which allows them to decay into $D+K$ mesons.
However, due to the threshold feature, the $DK$ decay channel in this state (2224.4MeV) is prohibited.
The two states with higher masses (2799.7MeV and 3041.6MeV) have larger amplitudes of the basis vector $\alpha_2$, which allows them to decay into $D^\ast+K^\ast$ mesons.
And the small amplitudes of the basis vector $\alpha_1$ mean that it can decay into $D+K$ mesons.
Therefore, our results show that the recently discovered $X_0(2900)$ state experimentally~\cite{Aaij:2020hon} is a good candidate for the $S$-wave tetraquark $nn\bar{s}\bar{c}$ state with the quantum number $0^+$.
And the $X_0(2900)$ state may be one of these two larger mass states. 
For 3041.6MeV state, it can also decay into $D^*+K^*$ mesons. The partial decay width ratio between two decay modes, namely $DK$ and $D^*K^*$ is,  
\begin{equation}
\frac{\Gamma[X(3041.6)\rightarrow D^*K^*]}{\Gamma[X(3041.6)\rightarrow DK]}=285.7.
\end{equation}
That means $D^*K^*$ is dominant decay mode. If we neglect the difference of initial states wavefunctions between 2543.5MeV, 2799.7MeV, and 3041.6MeV (that means with the same $\gamma_i$ in Eq.(\ref{dw})), the partial decay width ratio for $DK$ decay process can be expressed approximately,
\begin{eqnarray}
&&\Gamma[X(2543.5)\rightarrow D K]:\Gamma[X(2799.7)\rightarrow D K]:\nonumber\\
&&\Gamma[X(3041.6)\rightarrow D K] \approx 217.5:8.5:1.
\end{eqnarray}
Thus, The $X_0 (2900)$ state that has been discovered in the experiment is more likely the 2799.7MeV state in our calculation. This result is also consistent with the results in Refs.~\cite{Liu:2020nil,Huang:2020ptc,Agaev:2020nrc}. Besides, the state with mass 2543.5MeV which have very large partial decay width in $DK$ channel, it's most likely to be find in the future experiment.

For the quantum number $J^P=1^+$, the two states, 2417.9MeV and 2642.1MeV, have larger amplitudes of the basis vector $\beta_1$, which allows them to decay into $D^\ast+K$ mesons.
And due to the threshold feature, the $D^\ast+K$ decay progress of 2417.9MeV may be prohibited.
Similarly, the state with mass 2687.2MeV  can decay to $D^\ast+K$ mesons. 
The 2856MeV and 2866.3MeV can decay to $D^\ast+K$ or $D+K^\ast$ mesons.
The 2975.3MeV can decay to $D^\ast+K$, $D+K^\ast$ or $D^\ast+K^\ast$ mesons.
For the quantum number $J^P=2^+$, the masses of the possible tetraquark states are 2871.1MeV and 2965.7MeV, and the 2965.7MeV can decay easily to $D^\ast+K^\ast$ mesons because its mass is higher than the threshold of $D^\ast K^\ast$.
For the calculated results of the $ss\bar{n}\bar{c}$ system, we place them in Table~\ref{T1}(d) and Fig.~\ref{f1}(d).
The specific method is similar to the analysis of the structure of $nn\bar{s}\bar{c}$, so we will not discuss it in detail.
The partial decay width of each tetraquark state can be obtained by Eq.(\ref{dw}), which is also not calculated in detail here.

For tetraquark systems, $sn\bar{n}\bar{c}$ and $sn\bar{s}\bar{c}$, the calculated results are listed in Table~\ref{T1}(e), \ref{T1}(e$^\prime$), \ref{T1}(f) and \ref{T1}(f$^\prime$).
In addition, the relationship between their mass spectra and the threshold of the corresponding meson-meson is exhibited in Fig.~\ref{f1}(e) and Fig.~\ref{f1}(f).
Obviously, it can be seen that there are two kinds of different mesons decay channels in each of these two systems.
For example, the $sn\bar{n}\bar{c}$ system can decay into $(s\bar{n})(n\bar{c})$ channel or $(n\bar{n})(s\bar{c})$ channel.
Similarly, $sn\bar{s}\bar{c}$ can either decay into $(s\bar{s})(n\bar{c})$ channel or $(n\bar{s})(s\bar{c})$ channel.
Similar to the previous analysis method, we can know the quantum numbers and the decay modes of these possible tetraquark states in the $sn\bar{n}\bar{c}$ and $sn\bar{s}\bar{c}$ systems.
For these two systems, the corresponding decay channel analysis method is similar to the previous.

\subsection{Tetraquark states with $nn\bar{n}\bar{b}$,\, $ss\bar{s}\bar{b}$,\, $nn\bar{s}\bar{b}$,\, $ss\bar{n}\bar{b}$,\, $sn\bar{n}\bar{b}$,\, $sn\bar{s}\bar{b}$ configuration }

%---------------------------------------------------------------------------------------------------------------------
\begin{table*}[htbp]
\caption{\label{T2}The mass $M$ (in MeV) and the amplitudes of the basis vectors for the tetraquark $nn\bar{n}\bar{b}$, $ss\bar{s}\bar{b}$, $nn\bar{s}\bar{b}$, $ss\bar{n}\bar{b}$, $sn\bar{n}\bar{b}$ and $sn\bar{s}\bar{b}$ systems with the different quantum numbers $J^P=0^+$, $1^+$ and $2^+$.
The Hamiltonian and the basis have been introduced in the Sec.II.
The amplitudes $c_i$ in the (a), (b), (c), (d), (e), (f) parts correspond to the $|(q_1\bar{q}_3)(q_2\bar{q}_4)\rangle$ configure, while the amplitudes $c_i$ in the (e$^\prime$) and (f$^\prime$) parts correspond to the $|(q_1\bar{q}_4)(q_2\bar{q}_3)\rangle$ configure.}
\vspace{0.2cm}
\centering
\begin{tabular}{cccc|cccc}
 \hline \hline
System & $J^{P}$ & $M$(MeV) & $c_i$ &System & $J^{P}$ & $M$(MeV) & $c_i$     \\
 \hline\hline
$nn\bar{n}\bar{b}$
&$0^{+}$ & 6249.0& (0.04, -0.71, -0.69, -0.09) &        $ss\bar{s}\bar{b}$ &$0^{+}$ & 6519.6& (0.17, -0.64, -0.75, 0.05)\\
(a)&        & 6037.3& (-0.07, 0.63, -0.59, -0.50) &  (b)  &                & 6464.0& (0.02, 0.61, -0.56, -0.56) \\
&        & 5622.1& (-0.64, -0.27, 0.32, -0.64) &    &                      & 6253.5& (-0.62, -0.41, 0.17, -0.64) \\
&        & 5266.1& (-0.76, 0.14, -0.26, 0.58) &     &                      & 6082.9& (-0.76, 0.21, -0.33, 0.52) \\
&$1^{+}$ & 6225.3 & ({-0.05, -0.33, 0.62, 0.70, -0.14, 0.02})&   &$1^{+}$  & 6508.1& (-0.13, 0.10, 0.61, 0.54, -0.45, -0.33)\\
&        & 6113.6 &({-0.03, -0.67, -0.37, 0.10, 0.54, 0.33})&      &       & 6490.2& (0.19, 0.61, -0.15, -0.47, -0.58, -0.14)\\
&        & 6067.4 &({0.07, 0.07, -0.64, 0.48, -0.56, 0.21})&       &       & 6474.4& (-0.07, 0.10, -0.56, 0.59, -0.35, 0.45)\\
&        & 5986.3 &({0.01, 0.62, 0.09, 0.28, 0.43, 0.58})&         &       & 6337.7& (0.07, -0.70, -0.34, -0.10, -0.43, -0.45)\\
&        & 5675.8 &({0.60, -0.20, 0.25, -0.36, -0.32, 0.57})&      &       & 6284.0& (0.57, -0.31, 0.39, -0.13, -0.28, 0.58) \\
&        & 5341.1 &({-0.80, -0.08, 0.11, -0.26, -0.29, 0.43})&     &       & 6144.9& (-0.78, -0.16, 0.16, -0.35, -0.28, 0.36)\\
&$2^{+}$ & 6170.5         & (-0.58, 0.82)  &                       &$2^{+}$& 6533.8& (-0.58, 0.82)\\
&        & 6064.7        & (-0.82, -0.58)  &                       &       & 6385.4& (-0.82, -0.58) \\
\hline
$nn\bar{s}\bar{b}$
&$0^{+}$ & 6336.1& (-0.03, 0.72, 0.69, 0.10) &        $ss\bar{n}\bar{b}$ &$0^{+}$ & 6413.8& (0.14, -0.66, -0.74, 0.01)\\
(c)&        & 6129.0& (-0.10, 0.64, -0.60, -0.47) &   (d) &              & 6362.4& (-0.01, -0.61, 0.56, 0.56) \\
&        & 5926.9& (0.64, 0.26, -0.33, 0.64) &    &                      & 5991.2& (-0.63, -0.38, 0.21, -0.65) \\
&        & 5654.5& (-0.76, 0.11, -0.23, 0.60)  &    &                    & 5752.4& (-0.76, 0.20, -0.32, 0.52) \\
&$1^{+}$ & 6312.6 & (0.03, 0.32, -0.63, -0.69, 0.17, -0.02)&  &$1^{+}$   & 6399.7& (-0.12, 0.02, 0.63, 0.59, -0.41, -0.28)\\
&        & 6222.1 &(0.05, 0.69, 0.38, -0.14, -0.52, -0.30)&      &       & 6376.6& (-0.12, -0.66, 0.20, 0.30, 0.63, 0.17)\\
&        & 6168.1 &(0.12, -0.04, -0.62, 0.39, -0.66, 0.12)&      &       & 6370.0& (0.06, -0.04, 0.55, -0.64, 0.23, -0.49)\\
&        & 6098.2 &(0.04, 0.61, -0.03, 0.37, 0.34, 0.61)&        &       & 6203.0& (0.03, -0.69, -0.36, -0.09, -0.44, -0.43)\\
&        & 5978.4 &(0.56, -0.22, 0.27, -0.40, -0.26, 0.58)&      &       & 6026.2& (0.60, -0.27, 0.33, -0.19, -0.33, 0.56) \\
&        & 5730.7 &(-0.81, -0.07, 0.09, -0.23, -0.29, 0.43)&     &       & 5815.3& (-0.78, -0.14, 0.16, -0.33, -0.29, 0.39)\\
&$2^{+}$ & 6277.7         & (-0.58, 0.82)  &                     &$2^{+}$& 6420.9& (-0.58, 0.82)\\
&        & 6185.7        & (-0.82, -0.58)  &                     &       & 6250.2& (-0.82, -0.58) \\
%\hline
%$ns\bar{n}\bar{b}$
%&$0^{+}$ & 6336.3& (0.06, -0.56, -0.82, -0.12) &        $ns\bar{s}\bar{b}$ &$0^{+}$ & 6432.6& (0.055, -0.54, -0.83, -0.11)\\
%&        & 6191.3& (-0.04, 0.74, -0.44, -0.50) &    &                    & 6286.0& (0.05, -0.75, 0.43, 0.50) \\
%&        & 5860.2& (-0.33, -0.36, 0.34, -0.80) &    &                    & 6137.7& (-0.31, -0.38, 0.33, -0.81) \\
%&        & 5458.6& (-0.94, 0.06, -0.15, 0.30)  &   &                     & 5827.0& (-0.95, 0.05, -0.13, 0.29) \\
%&$1^{+}$ & 6315.3 & (-0.06, -0.21, 0.46, 0.83, -0.24, 0.01)&   &$1^{+}$  & 6412.6& (-0.06, -0.19, 0.45, 0.84, -0.26, -0.02)\\
%&        & 6245.3 &(-0.04, -0.55, -0.23, 0.18, 0.67, 0.39)&     &        & 6357.4& (0.07, 0.51, 0.19, -0.22, -0.74, -0.33)\\
%&        & 6207.3 &(-0.04, -0.19, 0.74, -0.34, 0.41, -0.37)&     &       & 6308.1& (0.08, 0.05, -0.74, 0.29, -0.43, 0.43)\\
%&        & 6098.0 &(-0.002, 0.73, 0.28, 0.14, 0.39, 0.47)&  &            & 6224.4& (-0.03, -0.72, -0.27, -0.12, -0.28, -0.56)\\
%&        & 5917.8 &(0.28, -0.28, 0.33, -0.36, -0.39, 0.67)&    &         & 6188.6& (0.24, -0.43, 0.38, -0.38, -0.33, 0.60) \\
%&        & 5517.4 &(0.96, 0.04, -0.05, 0.15, 0.15, -0.20)&   &           & 5885.0& (-0.96, -0.03, 0.03, -0.13, -0.14, 0.18)\\
%&$2^{+}$ & 6299.7         & (-0.50, 0.87)  &                     &$2^{+}$& 6411.4& (-0.45, 0.90)\\
%&        & 6155.1        & (-0.87, -0.50)  &                     &       & 6281.2& (-0.90, -0.45) \\
\hline
$sn\bar{n}\bar{b}$
&$0^{+}$ & 6336.3& (0.12, -0.80, -0.59, 0.04) &        $sn\bar{s}\bar{b}$ &$0^{+}$ & 6432.6& (0.14, -0.80, -0.58, 0.05)\\
(e)&        & 6191.3& (-0.02, 0.49, -0.69, -0.53) &   (f) &              & 6286.0& (-0.02, 0.48, -0.71, -0.52) \\
&        & 5860.2& (-0.87, -0.26, 0.15, -0.40) &    &                    & 6137.7& (-0.87, -0.27, 0.13, -0.39) \\
&        & 5458.6& (-0.49, 0.25, -0.39, 0.74)  &   &                     & 5827.0& (-0.47, 0.24, -0.38, 0.76) \\
&$1^{+}$ & 6314.5 &(-0.13, -0.32, 0.75, 0.55, -0.07, -0.13)&   &$1^{+}$  & 6412.4& (-0.16, -0.27, 0.77, 0.53, -0.07, -0.16)\\
&        & 6252.5 &(-0.07, -0.69, -0.26, 0.08, 0.60, 0.31)&     &        & 6367.1& (0.11, 0.72, 0.27, -0.15, -0.55, -0.27)\\
&        & 6207.7 &(0.005, 0.03, -0.47, 0.68, -0.44, 0.35)&     &        & 6305.9& (0.02, 0.007, -0.45, 0.67, -0.52, 0.29)\\
&        & 6090.8 &(0.04, -0.61, -0.25, -0.20, -0.51, -0.52)&  &         & 6214.4& (0.19, -0.61, -0.14, -0.27, -0.51, -0.49)\\
&        & 5899.0 &(-0.86, 0.17, -0.23, 0.14, 0.18, -0.37)&    &         & 6171.3& (-0.83, 0.11, -0.28, 0.07, 0.04, -0.46) \\
&        & 5536.6 &(-0.50, -0.16, 0.21, -0.41, -0.40, 0.60)&   &         & 5904.9& (-0.48, -0.17, 0.21, -0.42, -0.40, 0.60)\\
&$2^{+}$ & 6299.7         & (-0.65, 0.76)  &                     &$2^{+}$& 6411.4& (-0.70, 0.72)\\
&        & 6155.1        & (-0.76, -0.65)  &                     &       & 6281.2& (-0.72, -0.70) \\
\hline
$sn\bar{n}\bar{b}$
&$0^{+}$ & 6336.3& (-0.06, 0.56, 0.82, 0.11) &        $sn\bar{s}\bar{b}$ &$0^{+}$ & 6432.6& (-0.06, 0.54, 0.83, 0.10)\\
(e$^\prime$)&        & 6191.3& (-0.04, 0.74, -0.44, -0.50) &    (f$^\prime$) &     & 6286.0& (-0.05, 0.75,-0.43, -0.50) \\
&        & 5860.2& (0.33, 0.36, -0.34, 0.80) &    &                    & 6137.7& (0.31, 0.38, -0.33, 0.81) \\
&        & 5458.6& (-0.94, 0.06, -0.15, 0.29)  &   &                     & 5827.0& (-0.95, 0.05, -0.13, 0.29) \\
&$1^{+}$ & 6314.5 &(0.24, 0.06, 0.46, 0.23, -0.83, -0.03)&   &$1^{+}$  & 6412.4& (0.22, 0.06, 0.43, 0.27, -0.83, -0.07)\\
&        & 6252.5 &(0.34, 0.05, -0.20, -0.71, -0.22, 0.54)&     &        & 6367.1& (-0.30, -0.08, 0.13, 0.75, 0.26, -0.50)\\
&        & 6207.7 &(0.25, 0.001, 0.74, -0.42, 0.37, -0.28)&     &        & 6305.9& (0.16, -0.01, 0.80, -0.38, 0.35, -0.27)\\
&        & 6090.8 &(-0.83, -0.02, 0.36, -0.19, -0.11, 0.36)&  &          & 6214.4& (-0.80, -0.08, 0.34, -0.04, -0.09, 0.48)\\
&        & 5899.0 &(-0.27, 0.34, -0.27, -0.44, -0.31, -0.68)&    &       & 6171.3& (-0.44, 0.31, -0.20, -0.43, -0.28,-0.63)\\
&        & 5536.6 &(-0.04, -0.94, -0.08, -0.18, -0.18, -0.22)&   &       & 5904.9& (-0.04, -0.94, -0.09, -0.18, -0.17, -0.20)\\
&$2^{+}$ & 6299.7         & (-0.50, -0.87)  &                    &$2^{+}$& 6411.4          & (-0.45, -0.90)\\
&       & 6155.1          & (0.87, -0.50)  &                      &       & 6281.2          & (0.90, -0.45) \\
 \hline \hline
\end{tabular}
\end{table*}
%---------------------------------------------------------------------------------------------------------------------

%-----------------------------------------------------
\begin{figure*}[htbp]
\centering
\begin{minipage}{9.26cm}
    \includegraphics[width=0.9\textwidth,height=0.9\textwidth]{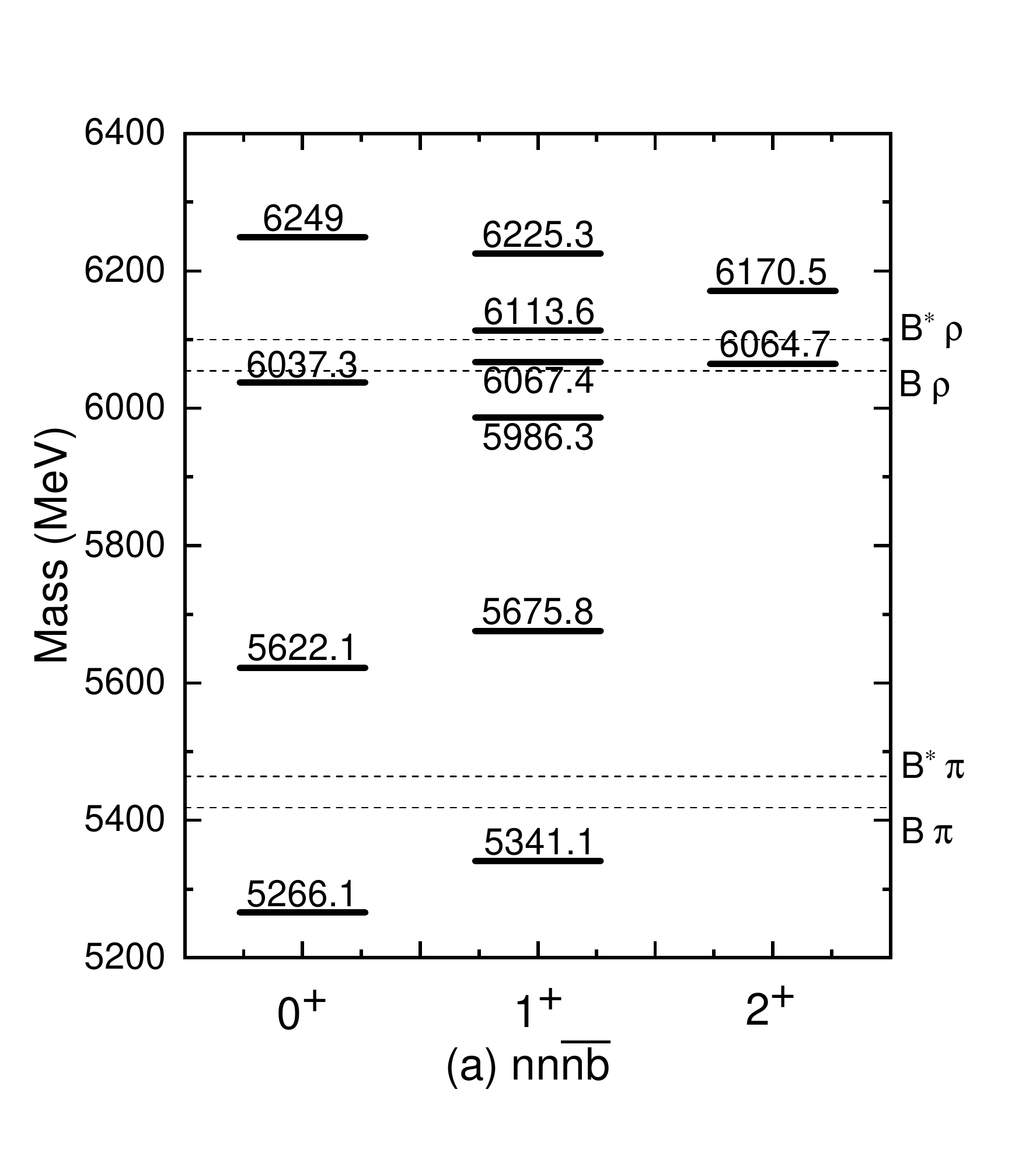}
  %\caption{}
  %\label{f2.12}
\end{minipage}
\hspace{-0.8cm}
\begin{minipage}{9.26cm}
    \includegraphics[width=0.9\textwidth,height=0.9\textwidth]{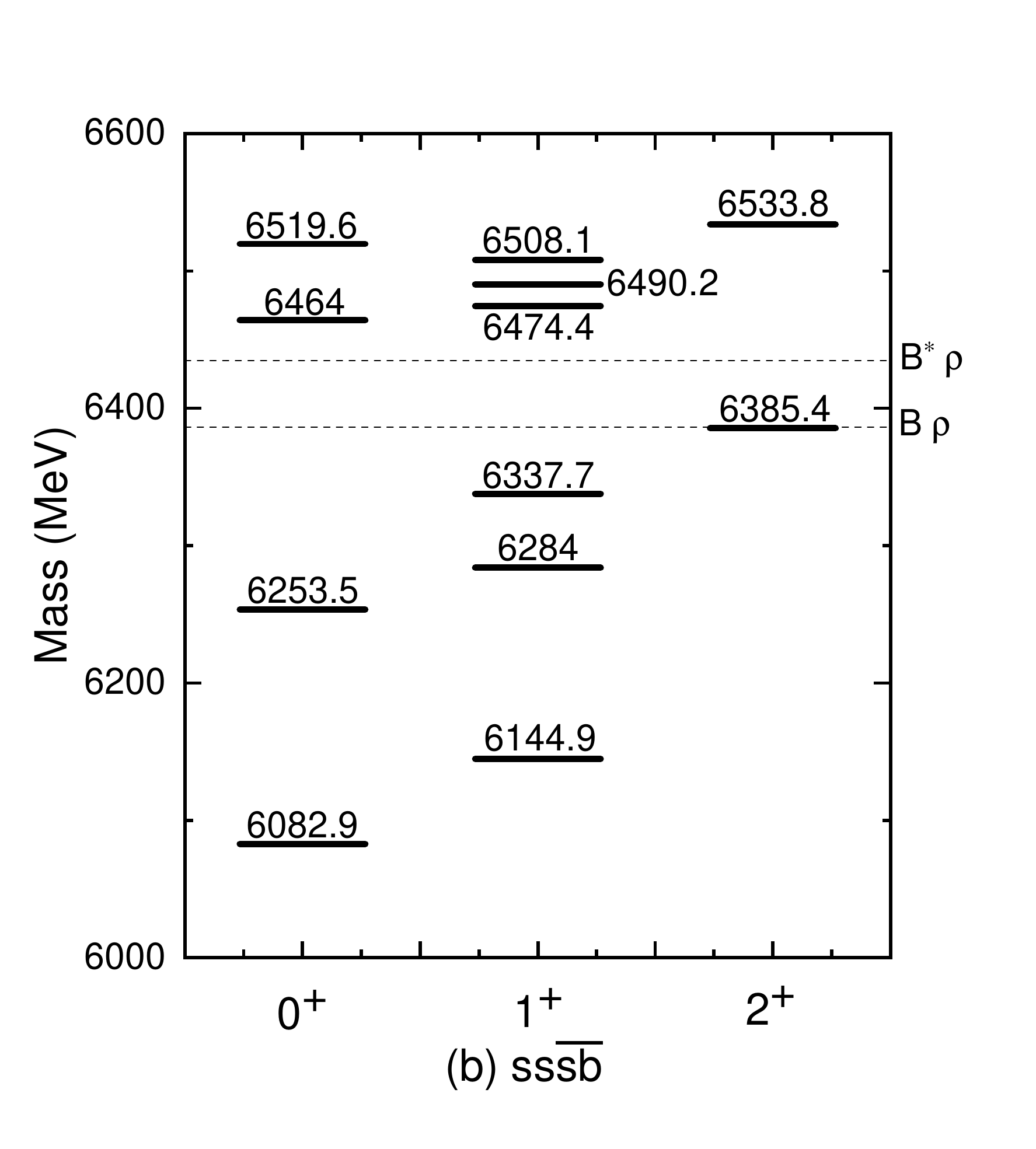}
\end{minipage}

\vspace{-1.2cm}

\begin{minipage}{9.26cm}
    \includegraphics[width=0.9\textwidth,height=0.9\textwidth]{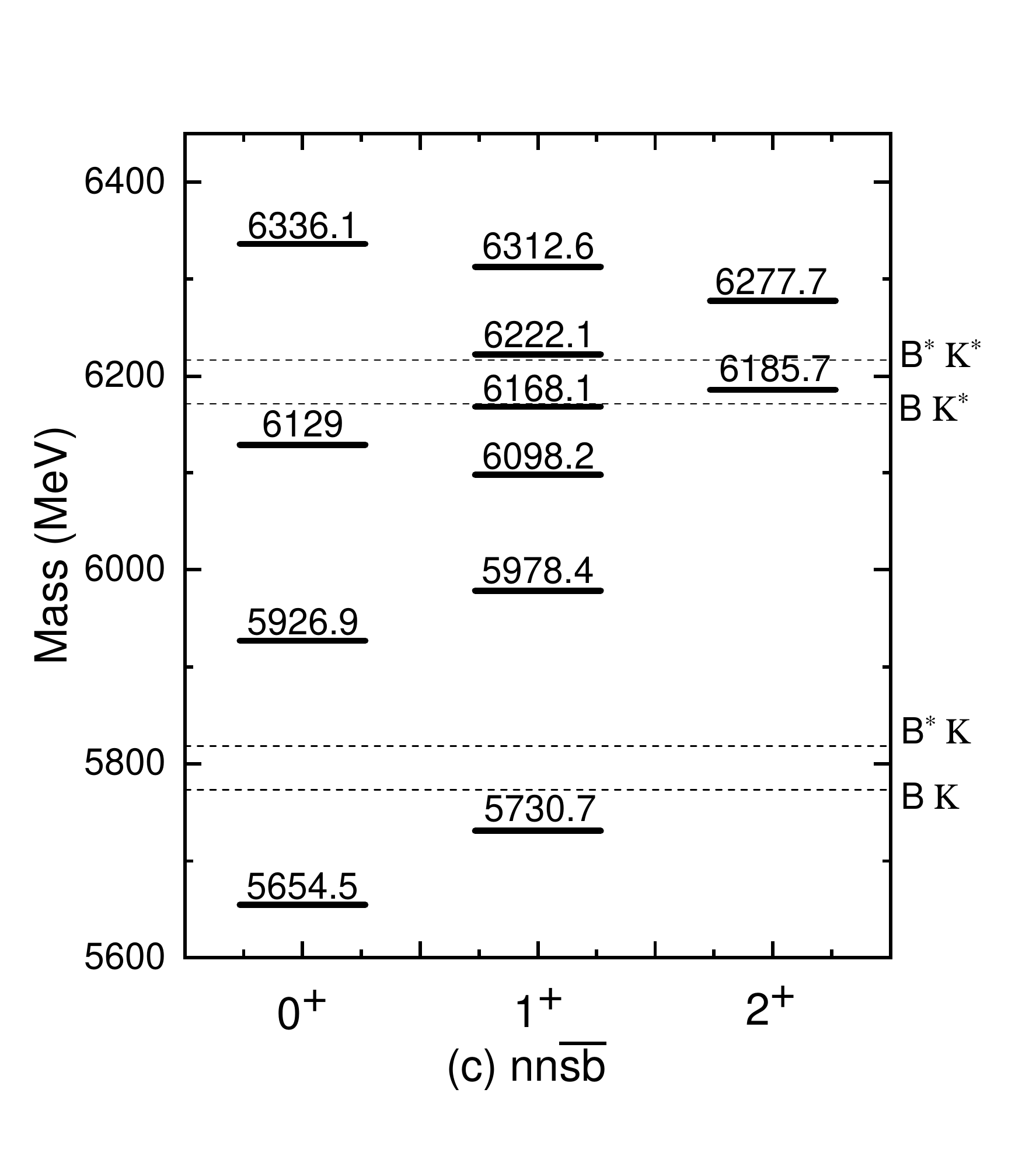}
\end{minipage}
\hspace{-0.8cm}
\begin{minipage}{9.26cm}
    \includegraphics[width=0.9\textwidth,height=0.9\textwidth]{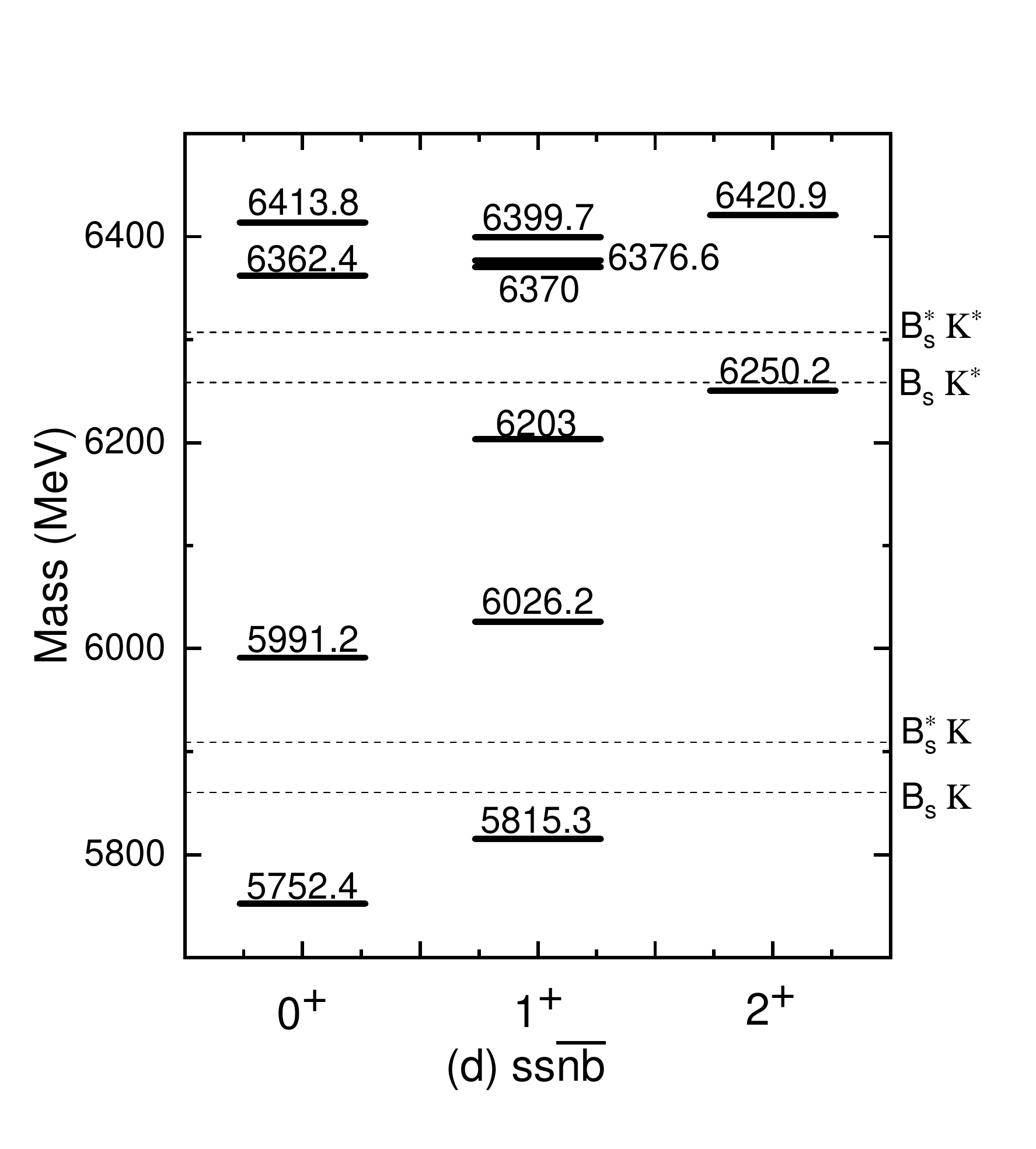}
\end{minipage}

\vspace{-1.2cm}

\begin{minipage}{9.26cm}
    \includegraphics[width=0.9\textwidth,height=0.9\textwidth]{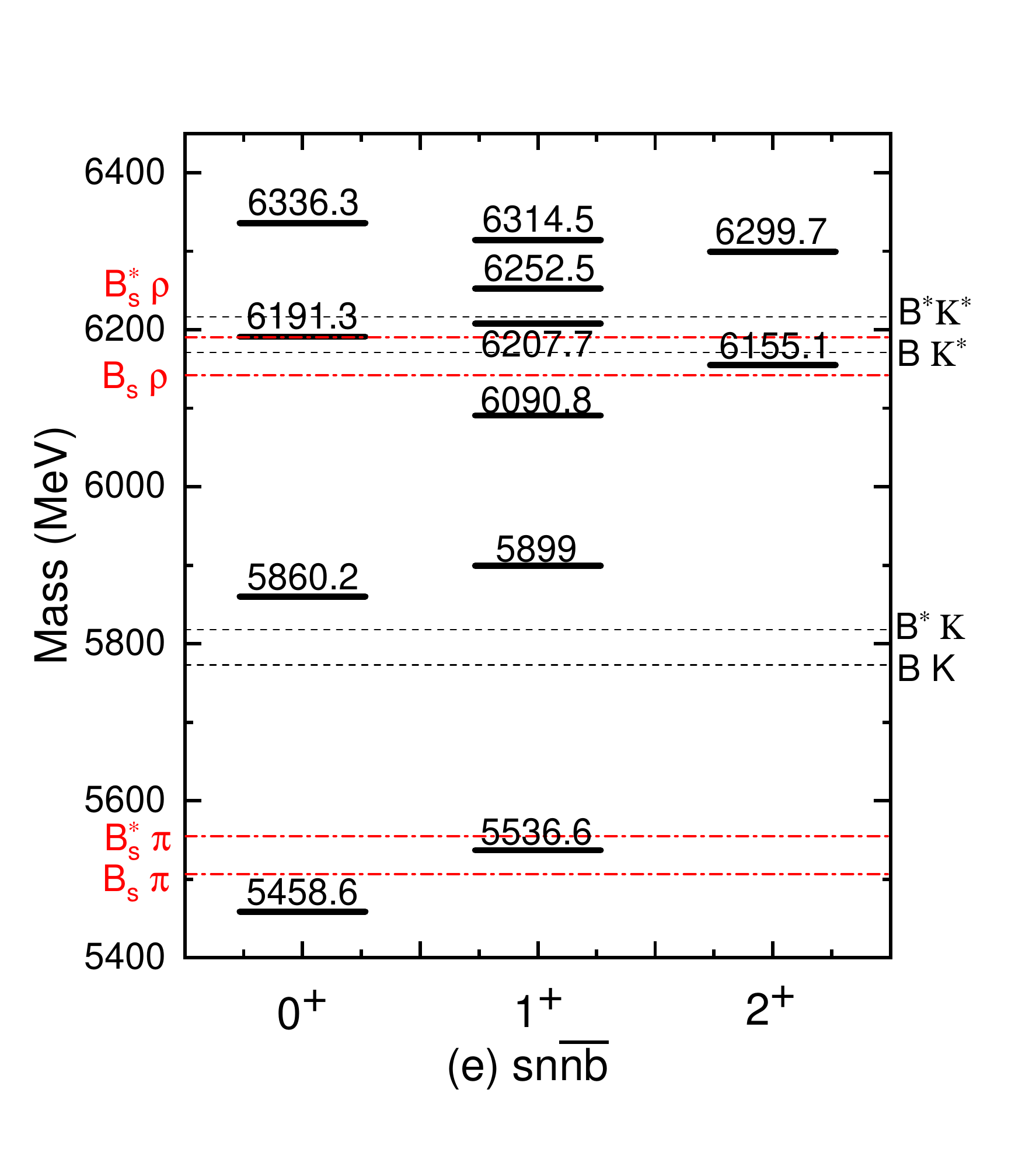}
\end{minipage}
\hspace{-0.8cm}
\begin{minipage}{9.26cm}
    \includegraphics[width=0.9\textwidth,height=0.9\textwidth]{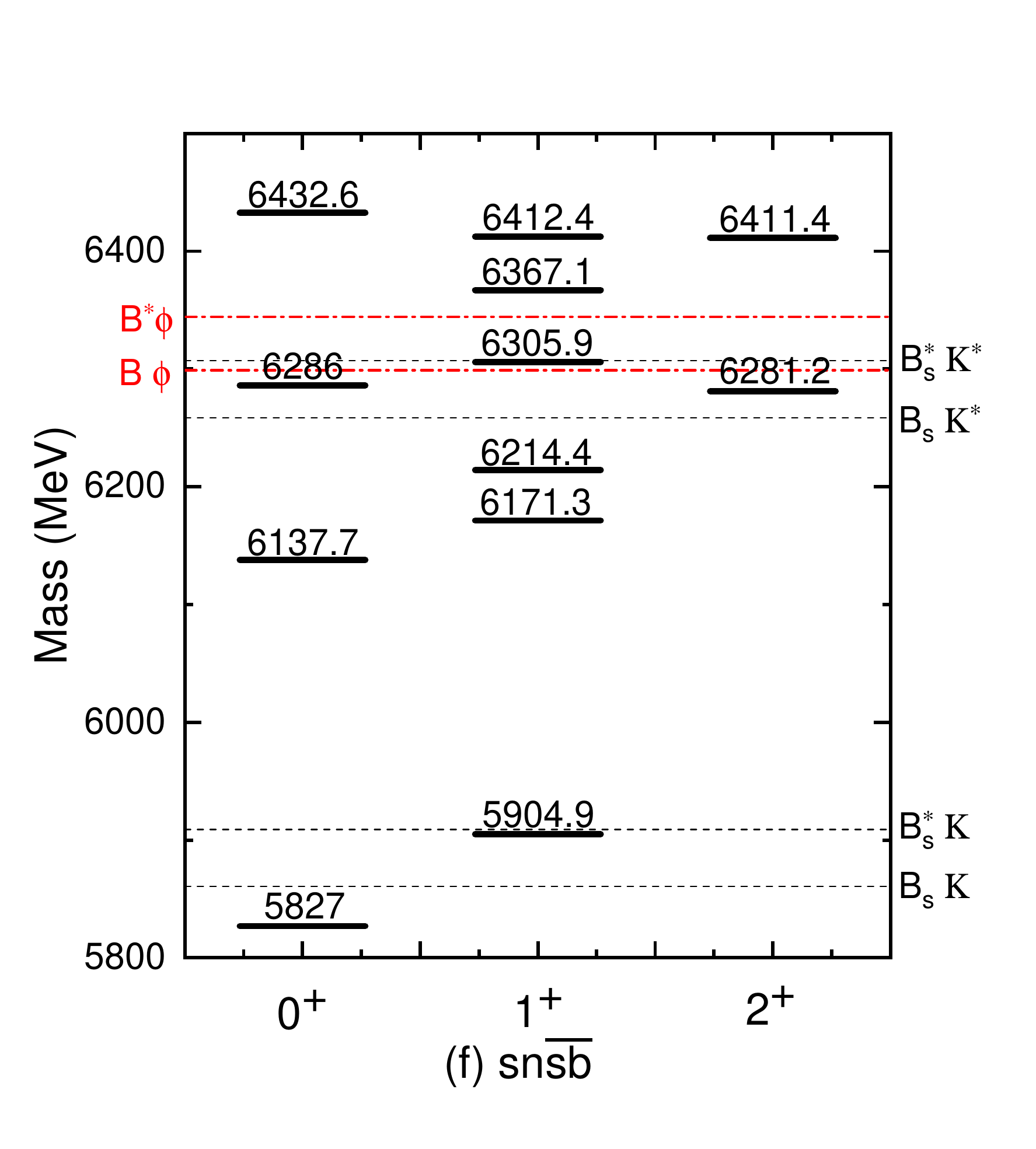}
\end{minipage}
\vspace{-0.5cm}
\caption{\label{f2}
The mass spectrum of the tetraquark system, $nn\bar{n}\bar{b}$, $ss\bar{s}\bar{b}$, $nn\bar{s}\bar{b}$, $ss\bar{n}\bar{b}$, $sn\bar{n}\bar{b}$ and $sn\bar{s}\bar{b}$.
The solid lines represent the mass spectrum of the possible tetraquark states.
The black dashed lines and the red dot-dashed lines are the thresholds of the possible meson-meson mass thresholds.
}
\end{figure*}
%-----------------------------------------------------

In this subsection, we further evaluated the mass spectra of the $S$-wave open bottom tetraquark systems.
Here the specific operation method is similar to the previous subsection.
Some required parameters in the calculation process have been extracted and shown in Table~\ref{mesonpm} to Table~\ref{baryonpv}.
Both the calculated mass spectra and the amplitudes of the color-spin wave function are exhibited in Table~\ref{T2}.
In order to better analyze the decay characteristics, we have drawn the possible tetraquark mass spectra and corresponding meson-meson thresholds in Fig.~\ref{f2}.
Here, we will firstly discuss the possible $S$-wave tetraquark states with four different flavors, such as the $nn\bar{s}\bar{b}$ systems with quantum number $0^+$ in detail.
There are four possible tetraquark states, and their masses are 5654.5MeV, 5926.9MeV, 6129.0MeV as well as 6336.1MeV in order. The state 5654.5MeV is lower than $BK$ threshold and decay is forbidden. 
The highest state 6336.1MeV which is high than $BK$ and $B^*K^*$ threshold, it can decay into both channel. The partial decay width ratio between this two decay modes gives,
\begin{equation}
	\frac{\Gamma[X(6336.1)\rightarrow B^*K^*]}{\Gamma[X(6336.1)\rightarrow B K]}=294.8.
\end{equation}
So, the dominant decay mode is the $B^*K^*$ final state. 
Exclude the lowest state, others have quite large amplitudes of the basis vector $\alpha_1$ and above the threshold, which allows them to decay into $D+K$ mesons. Similar with charm sector, it is probably to find $ud\bar{s}\bar{b}$ tetraquark state in the $BK$ decay channel with mass 5926.9MeV and 6129.0MeV.

Secondly, for the $sn\bar{n}\bar{b}$ and $sn\bar{s}\bar{b}$ systems,  any of them has two possible meson decay channels.
It can be seen that this $sn\bar{n}\bar{b}$ system has three possible quantum numbers $J^P=0^+$, $1^+$ and $2^+$ in Table~\ref{T2} and Fig.~\ref{f2}.
For the $sn\bar{n}\bar{b}$ system with the quantum number $0^+$, our results show that there are four possible tetraquark states, and their masses are 5458.2MeV, 5860.2MeV, 6191.3MeV, and 6336.3MeV, respectively.
The state, 5458.2MeV, have larger amplitudes of the basis vector $\alpha_1$.
It implies that the tetraquark state has a $B_s\pi$ or $BK$ decay channel.
But the mass spectrum is located below the $B_s\pi$ or $BK$ threshold.
Therefore, it is difficult to decay naturally into the $B_s+\pi$ or $B+K$ mesons.
Similarly, it can be concluded that 5860.2 MeV and 6191.3MeV can decay into the $B_s+\pi$ or $B+K$ mesons.
This highest state, 6191.3MeV, can decay into the $B_s+\pi$, $B+K$, $B_s^\ast + \rho$ or $B^*+K^*$ mesons.

For the $sn\bar{n}\bar{b}$ system with the quantum number $1^+$, the calculated results show that there are six possible $S$-wave tetraquark states. 
Their mass spectra and basis vector amplitudes are shown in Table~\ref{T2}(e) (or Table~\ref{T2}(e$^\prime$)) in turn.
Obviously, it can be seen that the lowest state 5536.6MeV has a very large amplitude, 0.94, in $B_s^\ast\pi$ decay channel.
But since the state is lower than the $B_s^\ast\pi$ threshold, it is difficult to decay naturally into a $B_s^\ast+\pi$ mesons.
In addition, the 5899MeV and 6090.8 can decay naturally into the $B_s^\ast+\pi$ or $B^\ast+K$ mesons.
The 6207.7MeV can decay naturally into the $B_s^\ast+\pi$, $B_s+\rho$, $B^\ast+K$ or $B+K^\ast$ mesons.
The 6252.5MeV and 6314.5MeV can decay naturally into the $B_s^\ast+\pi$, $B_s+\rho$, $B_s^\ast+\rho$, $B^\ast+K$, $B+K^\ast$ or $B^\ast+K^\ast$ mesons.

For the $sn\bar{n}\bar{b}$ system with the quantum number $2^+$, there are only two possible $S$-wave tetraquark states, and the masses are 6155.1MeV and 6299.7MeV respectively.
From both Fig.~\ref{f2}(e) and Table~\ref{T2}(e) (or Table~\ref{T2}(e$^\prime$)), it can be found that the dominant decay modes of these two states are $B^\ast K^\ast$ and $B_s^\ast \rho$ channel.
The corresponding larger mass state is easier to decay into $B^\ast K^\ast$ (or $B_s^\ast \rho$) because its mass is much larger than the $B^\ast K^\ast$ (or $B_s^\ast \rho$) threshold.
A state with a smaller mass is below the $B^\ast K^\ast$ (or $B_s^\ast \rho$) threshold, which implies that  it is difficult for this kind of decay to occur naturally.

For the $sn\bar{s}\bar{b}$ system with various quantum numbers, the calculated results are exhibited in Fig.~\ref{f2}(f) and Table~\ref{T2}(f) (or Table~\ref{T2}(f$^\prime$)).
For the quantum number $0^+$, our results indicate that the possible decay modes of the $sn\bar{s}\bar{b}$ states are as follows:
Although there is a $B_sK$ decay channel in the 5827MeV, it is difficult to naturally decay to $B_s+K$ mesons due to the small mass.
The main decay channel of the 6137.7MeV and 6286MeV is the $B_s+K$ mesons, and it can occur naturally.
This highest state, 6432.6MeV, can naturally decay into $B_s+K$, $B^\ast +\phi$ or $B_s^\ast+ K^\ast$ mesons.

For the quantum number $1^+$, the mass spectra and decay modes of the corresponding tetraquark states are as follows:
There is a $B^\ast_sK$ decay channel in the 5904.9MeV, but the mass is slightly smaller and cannot naturally decay into a $B^\ast_s+K$ mesons.
The decay mode of states 6171.3MeV and 6214.4MeV is mainly $B^\ast_sK$ channel.
The decay mode of the 6305.9MeV is mainly $B^\ast_sK$, $B_sK^\ast$ or $B\phi$ channel.
The decay modes of states 6367.1MeV and 6412.4MeV are mainly $B^\ast_sK$, $B_sK^\ast$, $B_s^\ast K^\ast$, $B\phi$ or $B^\ast\phi$ channel.
For the quantum number $2^+$, it is difficult for the 6281.2MeV to decay naturally into  $B_s^\ast + K^\ast$ or $B^\ast+\phi$ mesons because the mass is lower than the  $B_s^\ast K^\ast$ (or $B^\ast \phi$) threshold. 
But the 6411.4MeV can naturally decay into $B_s^\ast + K^\ast$ or $B^\ast+\phi$ mesons.
It is worth emphasizing that the corresponding partial decay widths of these above states can be obtained by Eq. (\ref{dw}), and no detailed calculations are made here.

In addition, properly considering the flavor symmetry breaking, we also calculated the mass spectra of the $nn\bar{n}\bar{b}$, $ss\bar{s}\bar{b}$, $nn\bar{s}\bar{b}$, and $ss\bar{n}\bar{b}$  systems as well as the amplitudes of the corresponding each color-spin wave function basis vector.
And all theoretical results are shown in Table~\ref{T2} and Fig.~\ref{f2}.
These will help us better understand the possible $S$-wave tetraquark states with single $b$ quark component.
The analysis of the properties of these possible tetraquark states are similar to the previous ones, and will not be elaborated here.

\section{summary}
In this paper, we perform a systematic study of the mass splitting of the $S$-wave tetraquark states containing the single heavy quark component ($c$ or $b$ quark) in the $SU(3)\otimes SU(2)$ color-spin space representations.
The method we adopt is an improved chromomagnetic interaction model, i.e., the chromoelectic interaction is properly considered in the conventional chromomagnetic interaction model. This is essential for studying the open heavy tetraquark states. First, we explain the mass spectra of the light and heavy flavor hadrons that have been discovered in the experiment.
Then, the color-spin wavefunctions of the tetraquark systems are constructed and the corresponding chromomagnetic and chromoelectic matrices are obtained.
By diagonalizing these matrices, we get the mass spectra and the amplitudes for different color-spin wavefunction bases of the possible tetraquark states.

For the low-lying $S$-wave tetraquark systems, all possible quantum numbers are assigned by $J^P=0^+$, $1^+$, and $2^+$. The mass splittings and the possible decay modes of all open heavy tetraquark states are calculated and exhibited in Table~\ref{T1} and Table~\ref{T2}.
Our calculation shows that $X_0(2900)$, which has been discovered recently, may be a good candidate for an $S$-wave tetraquark state of $ud\bar{s}\bar{c}$ with the quantum number of $0^+$. 
And in the same decay channel ($DK$), probably exists another $S$-wave tetraquark with a mass of around 2543.5MeV. In particular, the corresponding states in the bottom sector ($ud\bar{s}\bar{b}$) also exists with masses around 5926.9 and 6129.0MeV. We are looking forward to more theoretical researches and experiments in the future to check these possible tetraquark states.

\begin{acknowledgments}
The authors would like to thank Dr. G. Yang for helpful comments and discussions.
This work is supported by the NSFC under grant Nos. 11775123, 11890712, and 12047535, . 
\end{acknowledgments}

%-------------------------------------------------------------------------------------------------------------------------------------
%\nocite{*}
%\bibliographystyle{unsrt}
%\bibliographystyle{plain}
\bibliography{ref}

\providecommand{\noopsort}[1]{}\providecommand{\singleletter}[1]{#1}%
\begin{thebibliography}{74}
\expandafter\ifx\csname natexlab\endcsname\relax\def\natexlab#1{#1}\fi
\expandafter\ifx\csname bibnamefont\endcsname\relax
  \def\bibnamefont#1{#1}\fi
\expandafter\ifx\csname bibfnamefont\endcsname\relax
  \def\bibfnamefont#1{#1}\fi
\expandafter\ifx\csname citenamefont\endcsname\relax
  \def\citenamefont#1{#1}\fi
\expandafter\ifx\csname url\endcsname\relax
  \def\url#1{\texttt{#1}}\fi
\expandafter\ifx\csname urlprefix\endcsname\relax\def\urlprefix{URL }\fi
\providecommand{\bibinfo}[2]{#2}
\providecommand{\eprint}[2][]{\url{#2}}

\bibitem[{\citenamefont{Gell-Mann}(1964)}]{GellMann:1964nj}
\bibinfo{author}{\bibfnamefont{M.}~\bibnamefont{Gell-Mann}},
  \bibinfo{journal}{Phys. Lett.} \textbf{\bibinfo{volume}{8}},
  \bibinfo{pages}{214} (\bibinfo{year}{1964}).

\bibitem[{\citenamefont{Mathieu et~al.}(2009)\citenamefont{Mathieu, Kochelev,
  and Vento}}]{Mathieu:2008me}
\bibinfo{author}{\bibfnamefont{V.}~\bibnamefont{Mathieu}},
  \bibinfo{author}{\bibfnamefont{N.}~\bibnamefont{Kochelev}}, \bibnamefont{and}
  \bibinfo{author}{\bibfnamefont{V.}~\bibnamefont{Vento}},
  \bibinfo{journal}{Int. J. Mod. Phys. E} \textbf{\bibinfo{volume}{18}},
  \bibinfo{pages}{1} (\bibinfo{year}{2009}).

\bibitem[{\citenamefont{Celi et~al.}(2014)\citenamefont{Celi, Massignan,
  Ruseckas, Goldman, Spielman, Juzeli\={u}nas, and Lewenstein}}]{Celi:2013gma}
\bibinfo{author}{\bibfnamefont{A.}~\bibnamefont{Celi}},
  \bibinfo{author}{\bibfnamefont{P.}~\bibnamefont{Massignan}},
  \bibinfo{author}{\bibfnamefont{J.}~\bibnamefont{Ruseckas}},
  \bibinfo{author}{\bibfnamefont{N.}~\bibnamefont{Goldman}},
  \bibinfo{author}{\bibfnamefont{I.~B.} \bibnamefont{Spielman}},
  \bibinfo{author}{\bibfnamefont{G.}~\bibnamefont{Juzeli\={u}nas}},
  \bibnamefont{and}
  \bibinfo{author}{\bibfnamefont{M.}~\bibnamefont{Lewenstein}},
  \bibinfo{journal}{Phys. Rev. Lett.} \textbf{\bibinfo{volume}{112}},
  \bibinfo{pages}{043001} (\bibinfo{year}{2014}).

\bibitem[{\citenamefont{Esposito et~al.}(2017)\citenamefont{Esposito, Pilloni,
  and Polosa}}]{Esposito:2016noz}
\bibinfo{author}{\bibfnamefont{A.}~\bibnamefont{Esposito}},
  \bibinfo{author}{\bibfnamefont{A.}~\bibnamefont{Pilloni}}, \bibnamefont{and}
  \bibinfo{author}{\bibfnamefont{A.~D.} \bibnamefont{Polosa}},
  \bibinfo{journal}{Phys. Rept.} \textbf{\bibinfo{volume}{668}},
  \bibinfo{pages}{1} (\bibinfo{year}{2017}).

\bibitem[{\citenamefont{Karliner et~al.}(2018)\citenamefont{Karliner, Rosner,
  and Skwarnicki}}]{Karliner:2017qhf}
\bibinfo{author}{\bibfnamefont{M.}~\bibnamefont{Karliner}},
  \bibinfo{author}{\bibfnamefont{J.~L.} \bibnamefont{Rosner}},
  \bibnamefont{and}
  \bibinfo{author}{\bibfnamefont{T.}~\bibnamefont{Skwarnicki}},
  \bibinfo{journal}{Ann. Rev. Nucl. Part. Sci.} \textbf{\bibinfo{volume}{68}},
  \bibinfo{pages}{17} (\bibinfo{year}{2018}).

\bibitem[{\citenamefont{Meyer and Swanson}(2015)}]{Meyer:2015eta}
\bibinfo{author}{\bibfnamefont{C.~A.} \bibnamefont{Meyer}} \bibnamefont{and}
  \bibinfo{author}{\bibfnamefont{E.~S.} \bibnamefont{Swanson}},
  \bibinfo{journal}{Prog. Part. Nucl. Phys.} \textbf{\bibinfo{volume}{82}},
  \bibinfo{pages}{21} (\bibinfo{year}{2015}).

\bibitem[{\citenamefont{Chanowitz and Sharpe}(1983)}]{Chanowitz:1982qj}
\bibinfo{author}{\bibfnamefont{M.~S.} \bibnamefont{Chanowitz}}
  \bibnamefont{and} \bibinfo{author}{\bibfnamefont{S.~R.}
  \bibnamefont{Sharpe}}, \bibinfo{journal}{Nucl. Phys. B}
  \textbf{\bibinfo{volume}{222}}, \bibinfo{pages}{211} (\bibinfo{year}{1983}),
  \bibinfo{note}{[Erratum: Nucl.Phys.B 228, 588--588 (1983)]}.

\bibitem[{\citenamefont{De~Rujula et~al.}(1977)\citenamefont{De~Rujula, Georgi,
  and Glashow}}]{DeRujula:1976zlg}
\bibinfo{author}{\bibfnamefont{A.}~\bibnamefont{De~Rujula}},
  \bibinfo{author}{\bibfnamefont{H.}~\bibnamefont{Georgi}}, \bibnamefont{and}
  \bibinfo{author}{\bibfnamefont{S.~L.} \bibnamefont{Glashow}},
  \bibinfo{journal}{Phys. Rev. Lett.} \textbf{\bibinfo{volume}{38}},
  \bibinfo{pages}{317} (\bibinfo{year}{1977}).

\bibitem[{\citenamefont{Guo et~al.}(2018)\citenamefont{Guo, Hanhart,
  Mei\ss{}ner, Wang, Zhao, and Zou}}]{Guo:2017jvc}
\bibinfo{author}{\bibfnamefont{F.-K.} \bibnamefont{Guo}},
  \bibinfo{author}{\bibfnamefont{C.}~\bibnamefont{Hanhart}},
  \bibinfo{author}{\bibfnamefont{U.-G.} \bibnamefont{Mei\ss{}ner}},
  \bibinfo{author}{\bibfnamefont{Q.}~\bibnamefont{Wang}},
  \bibinfo{author}{\bibfnamefont{Q.}~\bibnamefont{Zhao}}, \bibnamefont{and}
  \bibinfo{author}{\bibfnamefont{B.-S.} \bibnamefont{Zou}},
  \bibinfo{journal}{Rev. Mod. Phys.} \textbf{\bibinfo{volume}{90}},
  \bibinfo{pages}{015004} (\bibinfo{year}{2018}).

\bibitem[{\citenamefont{Richard}(2016)}]{Richard:2016eis}
\bibinfo{author}{\bibfnamefont{J.-M.} \bibnamefont{Richard}},
  \bibinfo{journal}{Few Body Syst.} \textbf{\bibinfo{volume}{57}},
  \bibinfo{pages}{1185} (\bibinfo{year}{2016}).

\bibitem[{\citenamefont{Ali et~al.}(2017)\citenamefont{Ali, Lange, and
  Stone}}]{Ali:2017jda}
\bibinfo{author}{\bibfnamefont{A.}~\bibnamefont{Ali}},
  \bibinfo{author}{\bibfnamefont{J.~S.} \bibnamefont{Lange}}, \bibnamefont{and}
  \bibinfo{author}{\bibfnamefont{S.}~\bibnamefont{Stone}},
  \bibinfo{journal}{Prog. Part. Nucl. Phys.} \textbf{\bibinfo{volume}{97}},
  \bibinfo{pages}{123} (\bibinfo{year}{2017}).

\bibitem[{\citenamefont{Liu et~al.}(2019{\natexlab{a}})\citenamefont{Liu, Chen,
  Chen, Liu, and Zhu}}]{Liu:2019zoy}
\bibinfo{author}{\bibfnamefont{Y.-R.} \bibnamefont{Liu}},
  \bibinfo{author}{\bibfnamefont{H.-X.} \bibnamefont{Chen}},
  \bibinfo{author}{\bibfnamefont{W.}~\bibnamefont{Chen}},
  \bibinfo{author}{\bibfnamefont{X.}~\bibnamefont{Liu}}, \bibnamefont{and}
  \bibinfo{author}{\bibfnamefont{S.-L.} \bibnamefont{Zhu}},
  \bibinfo{journal}{Prog. Part. Nucl. Phys.} \textbf{\bibinfo{volume}{107}},
  \bibinfo{pages}{237} (\bibinfo{year}{2019}{\natexlab{a}}).

\bibitem[{\citenamefont{Brambilla et~al.}(2020)\citenamefont{Brambilla,
  Eidelman, Hanhart, Nefediev, Shen, Thomas, Vairo, and
  Yuan}}]{Brambilla:2019esw}
\bibinfo{author}{\bibfnamefont{N.}~\bibnamefont{Brambilla}},
  \bibinfo{author}{\bibfnamefont{S.}~\bibnamefont{Eidelman}},
  \bibinfo{author}{\bibfnamefont{C.}~\bibnamefont{Hanhart}},
  \bibinfo{author}{\bibfnamefont{A.}~\bibnamefont{Nefediev}},
  \bibinfo{author}{\bibfnamefont{C.-P.} \bibnamefont{Shen}},
  \bibinfo{author}{\bibfnamefont{C.~E.} \bibnamefont{Thomas}},
  \bibinfo{author}{\bibfnamefont{A.}~\bibnamefont{Vairo}}, \bibnamefont{and}
  \bibinfo{author}{\bibfnamefont{C.-Z.} \bibnamefont{Yuan}},
  \bibinfo{journal}{Phys. Rept.} \textbf{\bibinfo{volume}{873}},
  \bibinfo{pages}{1} (\bibinfo{year}{2020}).

\bibitem[{\citenamefont{Aaij et~al.}(2020{\natexlab{a}})}]{LHCb:2020bwg}
\bibinfo{author}{\bibfnamefont{R.}~\bibnamefont{Aaij}} \bibnamefont{et~al.}
  (\bibinfo{collaboration}{LHCb}), \bibinfo{journal}{Sci. Bull.}
  \textbf{\bibinfo{volume}{65}}, \bibinfo{pages}{1983}
  (\bibinfo{year}{2020}{\natexlab{a}}).

\bibitem[{\citenamefont{Karliner et~al.}(2017)\citenamefont{Karliner, Nussinov,
  and Rosner}}]{Karliner:2016zzc}
\bibinfo{author}{\bibfnamefont{M.}~\bibnamefont{Karliner}},
  \bibinfo{author}{\bibfnamefont{S.}~\bibnamefont{Nussinov}}, \bibnamefont{and}
  \bibinfo{author}{\bibfnamefont{J.~L.} \bibnamefont{Rosner}},
  \bibinfo{journal}{Phys. Rev. D} \textbf{\bibinfo{volume}{95}},
  \bibinfo{pages}{034011} (\bibinfo{year}{2017}).

\bibitem[{\citenamefont{L\"u et~al.}(2020)\citenamefont{L\"u, Chen, and
  Dong}}]{Lu:2020cns}
\bibinfo{author}{\bibfnamefont{Q.-F.} \bibnamefont{L\"u}},
  \bibinfo{author}{\bibfnamefont{D.-Y.} \bibnamefont{Chen}}, \bibnamefont{and}
  \bibinfo{author}{\bibfnamefont{Y.-B.} \bibnamefont{Dong}},
  \bibinfo{journal}{Eur. Phys. J. C} \textbf{\bibinfo{volume}{80}},
  \bibinfo{pages}{871} (\bibinfo{year}{2020}).

\bibitem[{\citenamefont{Liu et~al.}(2019{\natexlab{b}})\citenamefont{Liu, L\"u,
  Zhong, and Zhao}}]{Liu:2019zuc}
\bibinfo{author}{\bibfnamefont{M.-S.} \bibnamefont{Liu}},
  \bibinfo{author}{\bibfnamefont{Q.-F.} \bibnamefont{L\"u}},
  \bibinfo{author}{\bibfnamefont{X.-H.} \bibnamefont{Zhong}}, \bibnamefont{and}
  \bibinfo{author}{\bibfnamefont{Q.}~\bibnamefont{Zhao}},
  \bibinfo{journal}{Phys. Rev. D} \textbf{\bibinfo{volume}{100}},
  \bibinfo{pages}{016006} (\bibinfo{year}{2019}{\natexlab{b}}).

\bibitem[{\citenamefont{Wang et~al.}(2019)\citenamefont{Wang, Meng, and
  Zhu}}]{Wang:2019rdo}
\bibinfo{author}{\bibfnamefont{G.-J.} \bibnamefont{Wang}},
  \bibinfo{author}{\bibfnamefont{L.}~\bibnamefont{Meng}}, \bibnamefont{and}
  \bibinfo{author}{\bibfnamefont{S.-L.} \bibnamefont{Zhu}},
  \bibinfo{journal}{Phys. Rev. D} \textbf{\bibinfo{volume}{100}},
  \bibinfo{pages}{096013} (\bibinfo{year}{2019}).

\bibitem[{\citenamefont{Zhu}(2021)}]{Zhu:2020xni}
\bibinfo{author}{\bibfnamefont{R.}~\bibnamefont{Zhu}}, \bibinfo{journal}{Nucl.
  Phys. B} \textbf{\bibinfo{volume}{966}}, \bibinfo{pages}{115393}
  (\bibinfo{year}{2021}).

\bibitem[{\citenamefont{Bicudo et~al.}(2015)\citenamefont{Bicudo, Cichy,
  Peters, Wagenbach, and Wagner}}]{Bicudo:2015vta}
\bibinfo{author}{\bibfnamefont{P.}~\bibnamefont{Bicudo}},
  \bibinfo{author}{\bibfnamefont{K.}~\bibnamefont{Cichy}},
  \bibinfo{author}{\bibfnamefont{A.}~\bibnamefont{Peters}},
  \bibinfo{author}{\bibfnamefont{B.}~\bibnamefont{Wagenbach}},
  \bibnamefont{and} \bibinfo{author}{\bibfnamefont{M.}~\bibnamefont{Wagner}},
  \bibinfo{journal}{Phys. Rev. D} \textbf{\bibinfo{volume}{92}},
  \bibinfo{pages}{014507} (\bibinfo{year}{2015}).

\bibitem[{\citenamefont{Chen et~al.}(2017)\citenamefont{Chen, Chen, Liu,
  Steele, and Zhu}}]{Chen:2016jxd}
\bibinfo{author}{\bibfnamefont{W.}~\bibnamefont{Chen}},
  \bibinfo{author}{\bibfnamefont{H.-X.} \bibnamefont{Chen}},
  \bibinfo{author}{\bibfnamefont{X.}~\bibnamefont{Liu}},
  \bibinfo{author}{\bibfnamefont{T.~G.} \bibnamefont{Steele}},
  \bibnamefont{and} \bibinfo{author}{\bibfnamefont{S.-L.} \bibnamefont{Zhu}},
  \bibinfo{journal}{Phys. Lett. B} \textbf{\bibinfo{volume}{773}},
  \bibinfo{pages}{247} (\bibinfo{year}{2017}).

\bibitem[{\citenamefont{Zhao et~al.}(2020)\citenamefont{Zhao, Shi, and
  Zhuang}}]{Zhao:2020nwy}
\bibinfo{author}{\bibfnamefont{J.}~\bibnamefont{Zhao}},
  \bibinfo{author}{\bibfnamefont{S.}~\bibnamefont{Shi}}, \bibnamefont{and}
  \bibinfo{author}{\bibfnamefont{P.}~\bibnamefont{Zhuang}},
  \bibinfo{journal}{Phys. Rev. D} \textbf{\bibinfo{volume}{102}},
  \bibinfo{pages}{114001} (\bibinfo{year}{2020}), \eprint{2009.10319}.

\bibitem[{\citenamefont{Aaij et~al.}(2015)}]{LHCb:2015yax}
\bibinfo{author}{\bibfnamefont{R.}~\bibnamefont{Aaij}} \bibnamefont{et~al.}
  (\bibinfo{collaboration}{LHCb}), \bibinfo{journal}{Phys. Rev. Lett.}
  \textbf{\bibinfo{volume}{115}}, \bibinfo{pages}{072001}
  (\bibinfo{year}{2015}).

\bibitem[{\citenamefont{Aaij et~al.}(2019)}]{LHCb:2019kea}
\bibinfo{author}{\bibfnamefont{R.}~\bibnamefont{Aaij}} \bibnamefont{et~al.}
  (\bibinfo{collaboration}{LHCb}), \bibinfo{journal}{Phys. Rev. Lett.}
  \textbf{\bibinfo{volume}{122}}, \bibinfo{pages}{222001}
  (\bibinfo{year}{2019}).

\bibitem[{\citenamefont{Abazov et~al.}(2016)}]{D0:2016mwd}
\bibinfo{author}{\bibfnamefont{V.~M.} \bibnamefont{Abazov}}
  \bibnamefont{et~al.} (\bibinfo{collaboration}{D0}), \bibinfo{journal}{Phys.
  Rev. Lett.} \textbf{\bibinfo{volume}{117}}, \bibinfo{pages}{022003}
  (\bibinfo{year}{2016}).

\bibitem[{\citenamefont{Abazov et~al.}(2018)}]{D0:2017qqm}
\bibinfo{author}{\bibfnamefont{V.~M.} \bibnamefont{Abazov}}
  \bibnamefont{et~al.} (\bibinfo{collaboration}{D0}), \bibinfo{journal}{Phys.
  Rev. D} \textbf{\bibinfo{volume}{97}}, \bibinfo{pages}{092004}
  (\bibinfo{year}{2018}).

\bibitem[{\citenamefont{Aaij et~al.}(2016)}]{LHCb:2016dxl}
\bibinfo{author}{\bibfnamefont{R.}~\bibnamefont{Aaij}} \bibnamefont{et~al.}
  (\bibinfo{collaboration}{LHCb}), \bibinfo{journal}{Phys. Rev. Lett.}
  \textbf{\bibinfo{volume}{117}}, \bibinfo{pages}{152003}
  (\bibinfo{year}{2016}), \bibinfo{note}{[Addendum: Phys.Rev.Lett. 118, 109904
  (2017)]}.

\bibitem[{\citenamefont{Sirunyan et~al.}(2018)}]{CMS:2017hfy}
\bibinfo{author}{\bibfnamefont{A.~M.} \bibnamefont{Sirunyan}}
  \bibnamefont{et~al.} (\bibinfo{collaboration}{CMS}), \bibinfo{journal}{Phys.
  Rev. Lett.} \textbf{\bibinfo{volume}{120}}, \bibinfo{pages}{202005}
  (\bibinfo{year}{2018}).

\bibitem[{\citenamefont{Aaltonen et~al.}(2018)}]{CDF:2017dwr}
\bibinfo{author}{\bibfnamefont{T.}~\bibnamefont{Aaltonen}} \bibnamefont{et~al.}
  (\bibinfo{collaboration}{CDF}), \bibinfo{journal}{Phys. Rev. Lett.}
  \textbf{\bibinfo{volume}{120}}, \bibinfo{pages}{202006}
  (\bibinfo{year}{2018}).

\bibitem[{\citenamefont{Aaboud et~al.}(2018)}]{ATLAS:2018udc}
\bibinfo{author}{\bibfnamefont{M.}~\bibnamefont{Aaboud}} \bibnamefont{et~al.}
  (\bibinfo{collaboration}{ATLAS}), \bibinfo{journal}{Phys. Rev. Lett.}
  \textbf{\bibinfo{volume}{120}}, \bibinfo{pages}{202007}
  (\bibinfo{year}{2018}).

\bibitem[{\citenamefont{Aaij et~al.}(2020{\natexlab{b}})}]{LHCb:2020bls}
\bibinfo{author}{\bibfnamefont{R.}~\bibnamefont{Aaij}} \bibnamefont{et~al.}
  (\bibinfo{collaboration}{LHCb}), \bibinfo{journal}{Phys. Rev. Lett.}
  \textbf{\bibinfo{volume}{125}}, \bibinfo{pages}{242001}
  (\bibinfo{year}{2020}{\natexlab{b}}).

\bibitem[{\citenamefont{Aaij et~al.}(2020{\natexlab{c}})}]{LHCb:2020pxc}
\bibinfo{author}{\bibfnamefont{R.}~\bibnamefont{Aaij}} \bibnamefont{et~al.}
  (\bibinfo{collaboration}{LHCb}), \bibinfo{journal}{Phys. Rev. D}
  \textbf{\bibinfo{volume}{102}}, \bibinfo{pages}{112003}
  (\bibinfo{year}{2020}{\natexlab{c}}).

\bibitem[{\citenamefont{Molina et~al.}(2010)\citenamefont{Molina, Branz, and
  Oset}}]{Molina:2010tx}
\bibinfo{author}{\bibfnamefont{R.}~\bibnamefont{Molina}},
  \bibinfo{author}{\bibfnamefont{T.}~\bibnamefont{Branz}}, \bibnamefont{and}
  \bibinfo{author}{\bibfnamefont{E.}~\bibnamefont{Oset}},
  \bibinfo{journal}{Phys. Rev. D} \textbf{\bibinfo{volume}{82}},
  \bibinfo{pages}{014010} (\bibinfo{year}{2010}).

\bibitem[{\citenamefont{Chen et~al.}(2021)\citenamefont{Chen, Han, L\"u, Wang,
  and Yu}}]{Chen:2020eyu}
\bibinfo{author}{\bibfnamefont{Y.-K.} \bibnamefont{Chen}},
  \bibinfo{author}{\bibfnamefont{J.-J.} \bibnamefont{Han}},
  \bibinfo{author}{\bibfnamefont{Q.-F.} \bibnamefont{L\"u}},
  \bibinfo{author}{\bibfnamefont{J.-P.} \bibnamefont{Wang}}, \bibnamefont{and}
  \bibinfo{author}{\bibfnamefont{F.-S.} \bibnamefont{Yu}},
  \bibinfo{journal}{Eur. Phys. J. C} \textbf{\bibinfo{volume}{81}},
  \bibinfo{pages}{71} (\bibinfo{year}{2021}).

\bibitem[{\citenamefont{Guo et~al.}(2020)\citenamefont{Guo, Liu, and
  Sakai}}]{Guo:2019twa}
\bibinfo{author}{\bibfnamefont{F.-K.} \bibnamefont{Guo}},
  \bibinfo{author}{\bibfnamefont{X.-H.} \bibnamefont{Liu}}, \bibnamefont{and}
  \bibinfo{author}{\bibfnamefont{S.}~\bibnamefont{Sakai}},
  \bibinfo{journal}{Prog. Part. Nucl. Phys.} \textbf{\bibinfo{volume}{112}},
  \bibinfo{pages}{103757} (\bibinfo{year}{2020}).

\bibitem[{\citenamefont{Chen et~al.}(2020)\citenamefont{Chen, Chen, Dong, and
  Su}}]{Chen:2020aos}
\bibinfo{author}{\bibfnamefont{H.-X.} \bibnamefont{Chen}},
  \bibinfo{author}{\bibfnamefont{W.}~\bibnamefont{Chen}},
  \bibinfo{author}{\bibfnamefont{R.-R.} \bibnamefont{Dong}}, \bibnamefont{and}
  \bibinfo{author}{\bibfnamefont{N.}~\bibnamefont{Su}}, \bibinfo{journal}{Chin.
  Phys. Lett.} \textbf{\bibinfo{volume}{37}}, \bibinfo{pages}{101201}
  (\bibinfo{year}{2020}).

\bibitem[{\citenamefont{Agaev et~al.}(2020)\citenamefont{Agaev, Azizi, and
  Sundu}}]{Agaev:2020nrc}
\bibinfo{author}{\bibfnamefont{S.~S.} \bibnamefont{Agaev}},
  \bibinfo{author}{\bibfnamefont{K.}~\bibnamefont{Azizi}}, \bibnamefont{and}
  \bibinfo{author}{\bibfnamefont{H.}~\bibnamefont{Sundu}}
  (\bibinfo{year}{2020}), \eprint{2008.13027}.

\bibitem[{\citenamefont{Liu et~al.}(2020)\citenamefont{Liu, Xie, and
  Geng}}]{Liu:2020nil}
\bibinfo{author}{\bibfnamefont{M.-Z.} \bibnamefont{Liu}},
  \bibinfo{author}{\bibfnamefont{J.-J.} \bibnamefont{Xie}}, \bibnamefont{and}
  \bibinfo{author}{\bibfnamefont{L.-S.} \bibnamefont{Geng}},
  \bibinfo{journal}{Phys. Rev. D} \textbf{\bibinfo{volume}{102}},
  \bibinfo{pages}{091502} (\bibinfo{year}{2020}).

\bibitem[{\citenamefont{He and Chen}(2021)}]{He:2020btl}
\bibinfo{author}{\bibfnamefont{J.}~\bibnamefont{He}} \bibnamefont{and}
  \bibinfo{author}{\bibfnamefont{D.-Y.} \bibnamefont{Chen}},
  \bibinfo{journal}{Chin. Phys. C} \textbf{\bibinfo{volume}{45}},
  \bibinfo{pages}{063102} (\bibinfo{year}{2021}).

\bibitem[{\citenamefont{Molina and Oset}(2020)}]{Molina:2020hde}
\bibinfo{author}{\bibfnamefont{R.}~\bibnamefont{Molina}} \bibnamefont{and}
  \bibinfo{author}{\bibfnamefont{E.}~\bibnamefont{Oset}},
  \bibinfo{journal}{Phys. Lett. B} \textbf{\bibinfo{volume}{811}},
  \bibinfo{pages}{135870} (\bibinfo{year}{2020}).

\bibitem[{\citenamefont{Hu et~al.}(2021)\citenamefont{Hu, Lao, Ling, and
  Wang}}]{Hu:2020mxp}
\bibinfo{author}{\bibfnamefont{M.-W.} \bibnamefont{Hu}},
  \bibinfo{author}{\bibfnamefont{X.-Y.} \bibnamefont{Lao}},
  \bibinfo{author}{\bibfnamefont{P.}~\bibnamefont{Ling}}, \bibnamefont{and}
  \bibinfo{author}{\bibfnamefont{Q.}~\bibnamefont{Wang}},
  \bibinfo{journal}{Chin. Phys. C} \textbf{\bibinfo{volume}{45}},
  \bibinfo{pages}{021003} (\bibinfo{year}{2021}).

\bibitem[{\citenamefont{Wang}(2020)}]{Wang:2020xyc}
\bibinfo{author}{\bibfnamefont{Z.-G.} \bibnamefont{Wang}},
  \bibinfo{journal}{Int. J. Mod. Phys. A} \textbf{\bibinfo{volume}{35}},
  \bibinfo{pages}{2050187} (\bibinfo{year}{2020}).

\bibitem[{\citenamefont{Zhang}(2021)}]{Zhang:2020oze}
\bibinfo{author}{\bibfnamefont{J.-R.} \bibnamefont{Zhang}},
  \bibinfo{journal}{Phys. Rev. D} \textbf{\bibinfo{volume}{103}},
  \bibinfo{pages}{054019} (\bibinfo{year}{2021}).

\bibitem[{\citenamefont{Mutuk}(2021)}]{Mutuk:2020igv}
\bibinfo{author}{\bibfnamefont{H.}~\bibnamefont{Mutuk}}, \bibinfo{journal}{J.
  Phys. G} \textbf{\bibinfo{volume}{48}}, \bibinfo{pages}{055007}
  (\bibinfo{year}{2021}).

\bibitem[{\citenamefont{Wang et~al.}(2021)\citenamefont{Wang, Meng, Xiao, Oka,
  and Zhu}}]{Wang:2020prk}
\bibinfo{author}{\bibfnamefont{G.-J.} \bibnamefont{Wang}},
  \bibinfo{author}{\bibfnamefont{L.}~\bibnamefont{Meng}},
  \bibinfo{author}{\bibfnamefont{L.-Y.} \bibnamefont{Xiao}},
  \bibinfo{author}{\bibfnamefont{M.}~\bibnamefont{Oka}}, \bibnamefont{and}
  \bibinfo{author}{\bibfnamefont{S.-L.} \bibnamefont{Zhu}},
  \bibinfo{journal}{Eur. Phys. J. C} \textbf{\bibinfo{volume}{81}},
  \bibinfo{pages}{188} (\bibinfo{year}{2021}).

\bibitem[{\citenamefont{Yang et~al.}(2021)\citenamefont{Yang, Ping, and
  Segovia}}]{Yang:2021izl}
\bibinfo{author}{\bibfnamefont{G.}~\bibnamefont{Yang}},
  \bibinfo{author}{\bibfnamefont{J.}~\bibnamefont{Ping}}, \bibnamefont{and}
  \bibinfo{author}{\bibfnamefont{J.}~\bibnamefont{Segovia}},
  \bibinfo{journal}{Phys. Rev. D} \textbf{\bibinfo{volume}{103}},
  \bibinfo{pages}{074011} (\bibinfo{year}{2021}).

\bibitem[{\citenamefont{He et~al.}(2020)\citenamefont{He, Wang, and
  Zhu}}]{He:2020jna}
\bibinfo{author}{\bibfnamefont{X.-G.} \bibnamefont{He}},
  \bibinfo{author}{\bibfnamefont{W.}~\bibnamefont{Wang}}, \bibnamefont{and}
  \bibinfo{author}{\bibfnamefont{R.}~\bibnamefont{Zhu}}, \bibinfo{journal}{Eur.
  Phys. J. C} \textbf{\bibinfo{volume}{80}}, \bibinfo{pages}{1026}
  (\bibinfo{year}{2020}).

\bibitem[{\citenamefont{Silvestre-Brac}(1992)}]{Silvestre-Brac:1992kaa}
\bibinfo{author}{\bibfnamefont{B.}~\bibnamefont{Silvestre-Brac}},
  \bibinfo{journal}{Phys. Rev. D} \textbf{\bibinfo{volume}{46}},
  \bibinfo{pages}{2179} (\bibinfo{year}{1992}).

\bibitem[{\citenamefont{Buccella et~al.}(2007)\citenamefont{Buccella, Hogaasen,
  Richard, and Sorba}}]{Buccella:2006fn}
\bibinfo{author}{\bibfnamefont{F.}~\bibnamefont{Buccella}},
  \bibinfo{author}{\bibfnamefont{H.}~\bibnamefont{Hogaasen}},
  \bibinfo{author}{\bibfnamefont{J.-M.} \bibnamefont{Richard}},
  \bibnamefont{and} \bibinfo{author}{\bibfnamefont{P.}~\bibnamefont{Sorba}},
  \bibinfo{journal}{Eur. Phys. J. C} \textbf{\bibinfo{volume}{49}},
  \bibinfo{pages}{743} (\bibinfo{year}{2007}).

\bibitem[{\citenamefont{Hogaasen and Sorba}(2004)}]{Hogaasen:2004pm}
\bibinfo{author}{\bibfnamefont{H.}~\bibnamefont{Hogaasen}} \bibnamefont{and}
  \bibinfo{author}{\bibfnamefont{P.}~\bibnamefont{Sorba}},
  \bibinfo{journal}{Mod. Phys. Lett. A} \textbf{\bibinfo{volume}{19}},
  \bibinfo{pages}{2403} (\bibinfo{year}{2004}).

\bibitem[{\citenamefont{Anselmino et~al.}(1993)\citenamefont{Anselmino,
  Predazzi, Ekelin, Fredriksson, and Lichtenberg}}]{Anselmino:1992vg}
\bibinfo{author}{\bibfnamefont{M.}~\bibnamefont{Anselmino}},
  \bibinfo{author}{\bibfnamefont{E.}~\bibnamefont{Predazzi}},
  \bibinfo{author}{\bibfnamefont{S.}~\bibnamefont{Ekelin}},
  \bibinfo{author}{\bibfnamefont{S.}~\bibnamefont{Fredriksson}},
  \bibnamefont{and} \bibinfo{author}{\bibfnamefont{D.~B.}
  \bibnamefont{Lichtenberg}}, \bibinfo{journal}{Rev. Mod. Phys.}
  \textbf{\bibinfo{volume}{65}}, \bibinfo{pages}{1199} (\bibinfo{year}{1993}).

\bibitem[{\citenamefont{Lebed}(2015)}]{Lebed:2015tna}
\bibinfo{author}{\bibfnamefont{R.~F.} \bibnamefont{Lebed}},
  \bibinfo{journal}{Phys. Lett. B} \textbf{\bibinfo{volume}{749}},
  \bibinfo{pages}{454} (\bibinfo{year}{2015}).

\bibitem[{\citenamefont{Maiani et~al.}(2004)\citenamefont{Maiani, Piccinini,
  Polosa, and Riquer}}]{Maiani:2004uc}
\bibinfo{author}{\bibfnamefont{L.}~\bibnamefont{Maiani}},
  \bibinfo{author}{\bibfnamefont{F.}~\bibnamefont{Piccinini}},
  \bibinfo{author}{\bibfnamefont{A.~D.} \bibnamefont{Polosa}},
  \bibnamefont{and} \bibinfo{author}{\bibfnamefont{V.}~\bibnamefont{Riquer}},
  \bibinfo{journal}{Phys. Rev. Lett.} \textbf{\bibinfo{volume}{93}},
  \bibinfo{pages}{212002} (\bibinfo{year}{2004}).

\bibitem[{\citenamefont{Rossi and Veneziano}(2016)}]{Rossi:2016szw}
\bibinfo{author}{\bibfnamefont{G.}~\bibnamefont{Rossi}} \bibnamefont{and}
  \bibinfo{author}{\bibfnamefont{G.}~\bibnamefont{Veneziano}},
  \bibinfo{journal}{JHEP} \textbf{\bibinfo{volume}{06}}, \bibinfo{pages}{041}
  (\bibinfo{year}{2016}).

\bibitem[{\citenamefont{Luo et~al.}(2017)\citenamefont{Luo, Chen, Liu, Liu, and
  Zhu}}]{Luo:2017eub}
\bibinfo{author}{\bibfnamefont{S.-Q.} \bibnamefont{Luo}},
  \bibinfo{author}{\bibfnamefont{K.}~\bibnamefont{Chen}},
  \bibinfo{author}{\bibfnamefont{X.}~\bibnamefont{Liu}},
  \bibinfo{author}{\bibfnamefont{Y.-R.} \bibnamefont{Liu}}, \bibnamefont{and}
  \bibinfo{author}{\bibfnamefont{S.-L.} \bibnamefont{Zhu}},
  \bibinfo{journal}{Eur. Phys. J. C} \textbf{\bibinfo{volume}{77}},
  \bibinfo{pages}{709} (\bibinfo{year}{2017}).

\bibitem[{\citenamefont{Wu et~al.}(2017)\citenamefont{Wu, Liu, Chen, Liu, and
  Zhu}}]{Wu:2017weo}
\bibinfo{author}{\bibfnamefont{J.}~\bibnamefont{Wu}},
  \bibinfo{author}{\bibfnamefont{Y.-R.} \bibnamefont{Liu}},
  \bibinfo{author}{\bibfnamefont{K.}~\bibnamefont{Chen}},
  \bibinfo{author}{\bibfnamefont{X.}~\bibnamefont{Liu}}, \bibnamefont{and}
  \bibinfo{author}{\bibfnamefont{S.-L.} \bibnamefont{Zhu}},
  \bibinfo{journal}{Phys. Rev. D} \textbf{\bibinfo{volume}{95}},
  \bibinfo{pages}{034002} (\bibinfo{year}{2017}).

\bibitem[{\citenamefont{Wu et~al.}(2019)\citenamefont{Wu, Liu, Liu, and
  Zhu}}]{Wu:2018xdi}
\bibinfo{author}{\bibfnamefont{J.}~\bibnamefont{Wu}},
  \bibinfo{author}{\bibfnamefont{X.}~\bibnamefont{Liu}},
  \bibinfo{author}{\bibfnamefont{Y.-R.} \bibnamefont{Liu}}, \bibnamefont{and}
  \bibinfo{author}{\bibfnamefont{S.-L.} \bibnamefont{Zhu}},
  \bibinfo{journal}{Phys. Rev. D} \textbf{\bibinfo{volume}{99}},
  \bibinfo{pages}{014037} (\bibinfo{year}{2019}).

\bibitem[{\citenamefont{Cheng et~al.}(2020)\citenamefont{Cheng, Li, Liu, Liu,
  Si, and Yao}}]{Cheng:2020nho}
\bibinfo{author}{\bibfnamefont{J.-B.} \bibnamefont{Cheng}},
  \bibinfo{author}{\bibfnamefont{S.-Y.} \bibnamefont{Li}},
  \bibinfo{author}{\bibfnamefont{Y.-R.} \bibnamefont{Liu}},
  \bibinfo{author}{\bibfnamefont{Y.-N.} \bibnamefont{Liu}},
  \bibinfo{author}{\bibfnamefont{Z.-G.} \bibnamefont{Si}}, \bibnamefont{and}
  \bibinfo{author}{\bibfnamefont{T.}~\bibnamefont{Yao}},
  \bibinfo{journal}{Phys. Rev. D} \textbf{\bibinfo{volume}{101}},
  \bibinfo{pages}{114017} (\bibinfo{year}{2020}).

\bibitem[{\citenamefont{H\o{}gaasen et~al.}(2014)\citenamefont{H\o{}gaasen,
  Kou, Richard, and Sorba}}]{Hogaasen:2013nca}
\bibinfo{author}{\bibfnamefont{H.}~\bibnamefont{H\o{}gaasen}},
  \bibinfo{author}{\bibfnamefont{E.}~\bibnamefont{Kou}},
  \bibinfo{author}{\bibfnamefont{J.-M.} \bibnamefont{Richard}},
  \bibnamefont{and} \bibinfo{author}{\bibfnamefont{P.}~\bibnamefont{Sorba}},
  \bibinfo{journal}{Phys. Lett. B} \textbf{\bibinfo{volume}{732}},
  \bibinfo{pages}{97} (\bibinfo{year}{2014}).

\bibitem[{\citenamefont{Weng et~al.}(2018)\citenamefont{Weng, Chen, and
  Deng}}]{Weng:2018mmf}
\bibinfo{author}{\bibfnamefont{X.-Z.} \bibnamefont{Weng}},
  \bibinfo{author}{\bibfnamefont{X.-L.} \bibnamefont{Chen}}, \bibnamefont{and}
  \bibinfo{author}{\bibfnamefont{W.-Z.} \bibnamefont{Deng}},
  \bibinfo{journal}{Phys. Rev. D} \textbf{\bibinfo{volume}{97}},
  \bibinfo{pages}{054008} (\bibinfo{year}{2018}).

\bibitem[{\citenamefont{An et~al.}(2021)\citenamefont{An, Chen, Liu, and
  Liu}}]{An:2021vwi}
\bibinfo{author}{\bibfnamefont{H.-T.} \bibnamefont{An}},
  \bibinfo{author}{\bibfnamefont{K.}~\bibnamefont{Chen}},
  \bibinfo{author}{\bibfnamefont{Z.-W.} \bibnamefont{Liu}}, \bibnamefont{and}
  \bibinfo{author}{\bibfnamefont{X.}~\bibnamefont{Liu}},
  \bibinfo{journal}{Phys. Rev. D} \textbf{\bibinfo{volume}{103}},
  \bibinfo{pages}{114027} (\bibinfo{year}{2021}).

\bibitem[{\citenamefont{Weng et~al.}(2019)\citenamefont{Weng, Chen, Deng, and
  Zhu}}]{Weng:2019ynv}
\bibinfo{author}{\bibfnamefont{X.-Z.} \bibnamefont{Weng}},
  \bibinfo{author}{\bibfnamefont{X.-L.} \bibnamefont{Chen}},
  \bibinfo{author}{\bibfnamefont{W.-Z.} \bibnamefont{Deng}}, \bibnamefont{and}
  \bibinfo{author}{\bibfnamefont{S.-L.} \bibnamefont{Zhu}},
  \bibinfo{journal}{Phys. Rev. D} \textbf{\bibinfo{volume}{100}},
  \bibinfo{pages}{016014} (\bibinfo{year}{2019}).

\bibitem[{\citenamefont{De~Rujula et~al.}(1975)\citenamefont{De~Rujula, Georgi,
  and Glashow}}]{DeRujula:1975qlm}
\bibinfo{author}{\bibfnamefont{A.}~\bibnamefont{De~Rujula}},
  \bibinfo{author}{\bibfnamefont{H.}~\bibnamefont{Georgi}}, \bibnamefont{and}
  \bibinfo{author}{\bibfnamefont{S.~L.} \bibnamefont{Glashow}},
  \bibinfo{journal}{Phys. Rev. D} \textbf{\bibinfo{volume}{12}},
  \bibinfo{pages}{147} (\bibinfo{year}{1975}).

\bibitem[{\citenamefont{Maiani et~al.}(2005)\citenamefont{Maiani, Piccinini,
  Polosa, and Riquer}}]{Maiani:2004vq}
\bibinfo{author}{\bibfnamefont{L.}~\bibnamefont{Maiani}},
  \bibinfo{author}{\bibfnamefont{F.}~\bibnamefont{Piccinini}},
  \bibinfo{author}{\bibfnamefont{A.~D.} \bibnamefont{Polosa}},
  \bibnamefont{and} \bibinfo{author}{\bibfnamefont{V.}~\bibnamefont{Riquer}},
  \bibinfo{journal}{Phys. Rev. D} \textbf{\bibinfo{volume}{71}},
  \bibinfo{pages}{014028} (\bibinfo{year}{2005}).

\bibitem[{\citenamefont{Cui et~al.}(2006)\citenamefont{Cui, Chen, Deng, and
  Zhu}}]{Cui:2005az}
\bibinfo{author}{\bibfnamefont{Y.}~\bibnamefont{Cui}},
  \bibinfo{author}{\bibfnamefont{X.-L.} \bibnamefont{Chen}},
  \bibinfo{author}{\bibfnamefont{W.-Z.} \bibnamefont{Deng}}, \bibnamefont{and}
  \bibinfo{author}{\bibfnamefont{S.-L.} \bibnamefont{Zhu}},
  \bibinfo{journal}{Phys. Rev. D} \textbf{\bibinfo{volume}{73}},
  \bibinfo{pages}{014018} (\bibinfo{year}{2006}).

\bibitem[{\citenamefont{Hogaasen et~al.}(2006)\citenamefont{Hogaasen, Richard,
  and Sorba}}]{Hogaasen:2005jv}
\bibinfo{author}{\bibfnamefont{H.}~\bibnamefont{Hogaasen}},
  \bibinfo{author}{\bibfnamefont{J.~M.} \bibnamefont{Richard}},
  \bibnamefont{and} \bibinfo{author}{\bibfnamefont{P.}~\bibnamefont{Sorba}},
  \bibinfo{journal}{Phys. Rev. D} \textbf{\bibinfo{volume}{73}},
  \bibinfo{pages}{054013} (\bibinfo{year}{2006}).

\bibitem[{\citenamefont{Guo et~al.}(2011)\citenamefont{Guo, Cao, Zhou, and
  Chen}}]{Guo:2011gu}
\bibinfo{author}{\bibfnamefont{T.}~\bibnamefont{Guo}},
  \bibinfo{author}{\bibfnamefont{L.}~\bibnamefont{Cao}},
  \bibinfo{author}{\bibfnamefont{M.-Z.} \bibnamefont{Zhou}}, \bibnamefont{and}
  \bibinfo{author}{\bibfnamefont{H.}~\bibnamefont{Chen}}
  (\bibinfo{year}{2011}), \eprint{arXiv:1106.2284}.

\bibitem[{\citenamefont{Kim et~al.}(2015)\citenamefont{Kim, Cheoun, and
  Oh}}]{Kim:2014ywa}
\bibinfo{author}{\bibfnamefont{H.}~\bibnamefont{Kim}},
  \bibinfo{author}{\bibfnamefont{M.-K.} \bibnamefont{Cheoun}},
  \bibnamefont{and} \bibinfo{author}{\bibfnamefont{Y.}~\bibnamefont{Oh}},
  \bibinfo{journal}{Phys. Rev. D} \textbf{\bibinfo{volume}{91}},
  \bibinfo{pages}{014021} (\bibinfo{year}{2015}).

\bibitem[{\citenamefont{Oka and Takeuchi}(1989)}]{Oka:1989ud}
\bibinfo{author}{\bibfnamefont{M.}~\bibnamefont{Oka}} \bibnamefont{and}
  \bibinfo{author}{\bibfnamefont{S.}~\bibnamefont{Takeuchi}},
  \bibinfo{journal}{Phys. Rev. Lett.} \textbf{\bibinfo{volume}{63}},
  \bibinfo{pages}{1780} (\bibinfo{year}{1989}).

\bibitem[{\citenamefont{Zyla et~al.}(2020)}]{Zyla:2020zbs}
\bibinfo{author}{\bibfnamefont{P.~A.} \bibnamefont{Zyla}} \bibnamefont{et~al.}
  (\bibinfo{collaboration}{Particle Data Group}), \bibinfo{journal}{PTEP}
  \textbf{\bibinfo{volume}{2020}}, \bibinfo{pages}{083C01}
  (\bibinfo{year}{2020}).

\bibitem[{\citenamefont{Fey et~al.}(2019)\citenamefont{Fey, Yang, Rittenhouse,
  Munkes, Baluktsian, Schmelcher, Sadeghpour, and
  Shaffer}}]{PhysRevLett.122.103001}
\bibinfo{author}{\bibfnamefont{C.}~\bibnamefont{Fey}},
  \bibinfo{author}{\bibfnamefont{J.}~\bibnamefont{Yang}},
  \bibinfo{author}{\bibfnamefont{S.~T.} \bibnamefont{Rittenhouse}},
  \bibinfo{author}{\bibfnamefont{F.}~\bibnamefont{Munkes}},
  \bibinfo{author}{\bibfnamefont{M.}~\bibnamefont{Baluktsian}},
  \bibinfo{author}{\bibfnamefont{P.}~\bibnamefont{Schmelcher}},
  \bibinfo{author}{\bibfnamefont{H.~R.} \bibnamefont{Sadeghpour}},
  \bibnamefont{and} \bibinfo{author}{\bibfnamefont{J.~P.}
  \bibnamefont{Shaffer}}, \bibinfo{journal}{Phys. Rev. Lett.}
  \textbf{\bibinfo{volume}{122}}, \bibinfo{pages}{103001}
  (\bibinfo{year}{2019}).

\bibitem[{\citenamefont{Weng et~al.}(2021)\citenamefont{Weng, Chen, Deng, and
  Zhu}}]{Weng:2020jao}
\bibinfo{author}{\bibfnamefont{X.-Z.} \bibnamefont{Weng}},
  \bibinfo{author}{\bibfnamefont{X.-L.} \bibnamefont{Chen}},
  \bibinfo{author}{\bibfnamefont{W.-Z.} \bibnamefont{Deng}}, \bibnamefont{and}
  \bibinfo{author}{\bibfnamefont{S.-L.} \bibnamefont{Zhu}},
  \bibinfo{journal}{Phys. Rev. D} \textbf{\bibinfo{volume}{103}},
  \bibinfo{pages}{034001} (\bibinfo{year}{2021}).

\bibitem[{\citenamefont{Aaij et~al.}(2020{\natexlab{d}})}]{Aaij:2020hon}
\bibinfo{author}{\bibfnamefont{R.}~\bibnamefont{Aaij}} \bibnamefont{et~al.}
  (\bibinfo{collaboration}{LHCb}), \bibinfo{journal}{Phys. Rev. Lett.}
  \textbf{\bibinfo{volume}{125}}, \bibinfo{pages}{242001}
  (\bibinfo{year}{2020}{\natexlab{d}}).

\bibitem[{\citenamefont{Huang et~al.}(2020)\citenamefont{Huang, Lu, Xie, and
  Geng}}]{Huang:2020ptc}
\bibinfo{author}{\bibfnamefont{Y.}~\bibnamefont{Huang}},
  \bibinfo{author}{\bibfnamefont{J.-X.} \bibnamefont{Lu}},
  \bibinfo{author}{\bibfnamefont{J.-J.} \bibnamefont{Xie}}, \bibnamefont{and}
  \bibinfo{author}{\bibfnamefont{L.-S.} \bibnamefont{Geng}},
  \bibinfo{journal}{Eur. Phys. J. C} \textbf{\bibinfo{volume}{80}},
  \bibinfo{pages}{973} (\bibinfo{year}{2020}).

\end{thebibliography}

%\begin{thebibliography}{99}
%\end{thebibliography}
%--------------------------------------------------------------------------------------------------------------------------------

\end{document}